# WFABC: a Wright-Fisher ABC-based approach for inferring effective population sizes and selection coefficients from time-sampled data

Matthieu Foll[*,§,1,2], Hyunjin Shim[*,1,2], and Jeffrey D. Jensen[1,2]

[*] authors contributed equally

[§] corresponding author

[1]School of Life Sciences, Ecole Polytechnique Fédérale de Lausanne (EPFL), Lausanne, Switzerland

[2]Swiss Institute of Bioinformatics (SIB), Lausanne, Switzerland

**Address for all co-authors:**

EPFL SV IBI-SV UPJENSEN

AAB 0 43 (Bâtiment AAB)

Station 15

CH-1015 Lausanne

SWITZERLAND

**Corresponding author:** Matthieu Foll (matthieu.foll@epfl.ch)

**Running title:** ABC method for time-sampled data

**Abstract**

With novel developments in sequencing technologies, time-sampled data are becoming more available and accessible. Naturally, there have been efforts in parallel to infer population genetic parameters from these datasets. Here, we compare and analyze four recent approaches based on the Wright-Fisher model for inferring selection coefficients ($s$) given effective population size ($N_e$), with simulated temporal datasets. Furthermore, we demonstrate the advantage of a recently proposed ABC-based method that is able to correctly infer genome-wide average $N_e$ from time-serial data, which is then set as a prior for inferring per-site selection coefficients accurately and precisely. We implement this ABC method in a new software and apply it to a classical time-serial dataset of the *medionigra* genotype in the moth *Panaxia dominula*. We show that a recessive lethal model is the best explanation for the observed variation in allele frequency by implementing an estimator of the dominance ratio ($h$).

**Introduction**

The study of temporal changes in allele frequency originated with two of the early founders of population genetics (Fisher 1922; Wright 1931), in which the fate of an allele was considered under a variety of models – including neutrality, positive and negative selection, and migration. Their celebrated debate on the relative roles of selection and drift in shaping the course of evolution also encompassed time-sampled data (Fisher and Ford 1947; Wright 1948), upon the publication of the time-series analysis of the *medionigra* phenotype in the moth *Panaxia dominula*. Following on this, and alternatively taking an experimental evolution approach, Clegg studied the dynamics of gene frequency change in *Drosophila melanogaster* (Clegg *et al.* 1976;

Cavener and Clegg 1978; Clegg 1978). However, owing to the limited availability of genetics markers, relatively few time-sampled datasets were available for consideration throughout the remainder of the 20th century. Thus, most test statistics for distinguishing the effects of selection from drift were focused on single time-point datasets – basing inference on patterns in the site frequency spectrum, linkage disequilibrium, and polymorphism/divergence (for a review, see Crisci *et al.* 2012).

Recently, sequencing data from multiple time-points has become increasingly common owing to novel developments in sequencing technologies (Schuster 2007) – coming from the fields of both ancient genomics and experimental evolution. This additional temporal component has the promise of providing improved power for inferring population genetic parameters compared to single time-point based analyses, as the trajectory of the allele itself provides valuable information about the underlying selection coefficient.

However, there are a limited number of methods currently available to estimate these parameters from time-sampled data. Moment-based methods (Kimura and Crow 1963; Pamilo and Varvio-Aho 1980; Nei and Tajima 1981; Waples 1989; Jorde and Ryman 2007) have been proposed utilizing the variance of gene frequency changes to infer effective population size ($N_e$). In addition, likelihood-based methods (Williamson and Slatkin 1999; Anderson *et al.* 2000; Berthier *et al.* 2002; Anderson 2005) have been proposed to calculate the probability of a given data observation given a pre-defined model. Efforts to incorporate selection into these estimation procedures have only recently begun, and given the rapidly increasing availability of such sequencing datasets, we now have a unique opportunity to re-address the puzzle of distinguishing genetic drift from selection with greater precision and power.

Thus, we present here new software implementing and expanding an approximate Bayesian computation (ABC, Sunnåker *et al.* 2013) approach to jointly infer per-site effective population sizes ($N_e$) and selection coefficients (*s*) from time-sampled data, initially described in Foll *et al.* (2014) to search for resistance mutations in time-sampled data from the influenza virus. Furthermore, we compare this approach with existing likelihood-based methods (Bollback *et al.* 2008; Malaspinas *et al.* 2012; Mathieson and McVean 2013), in order to inform future users on the most suitable method to be applied for any given dataset.

## Materials and Methods

### $N_e$-based ABC method

The data *X* consists of allele frequency trajectories measured at *L* loci: $x_i (i = 1,...,L)$. The $N_e$-based ABC methodology infers both the effective population size $N_e$ shared by all loci, and *L* locus-specific selection coefficients $s_i (i = 1,...,L)$. At a particular locus *i*, we can approximate the joint posterior distribution as (see Foll *et al.* 2014 for details):

$$P(N_e, s_i | X) \approx P(N_e | T(X)) P(s_i | N_e, U(X_i))$$

where $T(X) = T(X_1,...,X_L)$ denotes summary statistics that are a function of all loci together chosen to be informative about $N_e$, and *U(Xi)* denotes locus-specific summary statistics chosen to be informative about *si*. A two-step ABC algorithm as proposed by Bazin *et al.* (2010) is used to approximate this posterior:

Step 1. Obtain an approximation of the density

$$P(N_e | T(X)) \approx P(N_e | X)$$

Step 2. For locus *i=1* to *i=L*:

i. Simulate *K* trajectories $X_{i,k}$ from a Wright-Fisher model with $s_i$ randomly sampled from its prior and $N_e$ from the density obtained in step 1.

ii. Compute $U(X_{i,k})$ for each simulated trajectory.

iii. Retain the simulations with the smallest Euclidian distance between $U(X_i)$ and $U(x_i)$ to obtain a sample from an approximation to

$$P(s_i | N_e, X_i) P(N_e | X) = P(N_e, s_i | X).$$

In the original algorithm (Bazin *et al.* 2010), the first step is also achieved using ABC. In our case, we define *T(X)* as a single statistic given by Jorde and Ryman (2007) *Fs'* unbiased estimator of $N_e$:

$$Fs' = \frac{1}{t} \frac{Fs\left[1 - 1/(2\tilde{n})\right] - 2/\tilde{n}}{(1 + Fs/4)\left[1 - 1/(n_y)\right]} \text{ with } Fs = \frac{(x-y)^2}{z(1-z)}$$

where *x* and *y* are the minor allele frequencies at the two time points separated by *t* generations, *z=(x+y)/2*, and $\tilde{n}$ is the harmonic mean of the sample sizes $n_x$ and $n_y$ at the two time points expressed in number of chromosomes (twice the number of individuals for diploids). We average *Fs'* values over sites and times to obtain a genome wide estimator of $N_e=1/Fs'$ for haploids and $N_e=1/2Fs'$ for diploids (Jorde and Ryman 2007). A Bayesian bootstrap approach (Rubin 1981) is used to obtain a distribution for $P(N_e | T(X))$. Please note that we use the common notation where the effective population size $N_e$ corresponds to number of individuals, and the corresponding number of chromosomes for diploids is $2N_e$.

In the second step, simulations are performed using a Wright-Fisher model with an initial allele frequency and sample sizes matching the observed (simulation code available in the downloadable software package). At each site, we utilize two summary statistics derived from *Fs'*: $U(X_i)= (Fsd'_i, Fsi'_i)$ with *Fsd'* and *Fsi'* calculated respectively between pairs of time points where the allele considered is decreasing and increasing in frequency, such that at a given site $Fs'=Fsd' + Fsi'$. For the diploid model, we define the relative fitness as $w_{AA}=1+s$, $w_{Aa}=1+sh$ and $w_{aa}=1$ where $h$ denotes the dominance ratio (1 = dominant, 0.5 = co-dominance, 0 = recessive); and as $w_A=1+s$ and $w_a=1$ for the haploid model (Ewens, 2004).

We here implement this two-step approach in a new command line C++ program termed 'Wright-Fisher ABC' (WFABC). This estimation procedure is suitable for both haploid and diploid models of selection. Source code and binary executables for Linux, OS X, and Windows are freely available from the 'software' page of the Jensen Lab website: http://jensenlab.epfl.ch/

**Likelihood-based methods**

Currently, there are three likelihood-based methods available for inferring population genetic parameters from time-serial data (Bollback *et al.* 2008; Malaspinas *et al.* 2012; Mathieson and McVean 2013), based on a Hidden Markov Model (HMM) to model the allele frequency trajectory.

Bollback *et al.* (2008) co-estimates the selection coefficient ($s$) and the effective population size ($N_e$) from a diffusion process, by approximating the Wright-Fisher model and computing the maximum likelihood at fixed intervals. Malaspinas *et al.* (2012) additionally estimates the allele age ($t_0$), and further approximates the Wright-Fisher model through a one-step process. Mathieson and McVean (2013) estimates only the selection coefficient ($s$) assuming that $N_e$ is known, using an expectation-maximization (EM) algorithm that can be extended to the case of a structured population. The fitness is parameterized as with our $N_e$-based ABC approach (see above). The likelihood function of the parameters of interest - $\theta=(\gamma,N_e)$ for the Bollback *et al.* (2008) method and $\theta=(\gamma,N_e,t_0)$ for the Malaspinas *et al.* (2012) method, where $\gamma=2N_es$ – is conditioned over all population allelic frequencies $x_{j_1},...,x_{j_m}$ at sampling times $T=(t_1,...,t_m)$, and is given as:

$$l(\theta) = \sum_{j_1}...\sum_{j_m} p(i_1,...,i_m \mid \theta, T, x_{j_1},...,x_{j_m}) p(x_{j_1},...,x_{j_m} \mid \theta, T)$$

where $i_k$ is the frequency of the minor allele at the sampling time $k$, and $j_k$ is the true minor allele frequency at the sampling time $k$, where $x_{j_k} = j_k / 2N_e$. The first term of the likelihood is the emission probability, which is modeled as a binomial sampling, and the second term represents the transition probabilities in the Markov chain. In both methods the Markov chain is approximated by a diffusion process, from which the transition probabilities are given as the backward Kolmogorov equations (Ewens, 2004).

    The major difference between these three likelihood-based methods comes from the implementation of how these probabilities are calculated. The Bollback *et al.* (2008) method utilizes numerical approximations to evaluate the likelihood function, first by using the Crank-Nicolson approximation (Crank *et al.* 1947) for the backward Kolmogorov equation, and second by using numerical integration for the emission probability. Mathiesion and McVean (2013) use an expectation-maximization (EM) algorithm to find the maximum likelihood estimate (MLE) of *s* based on the MLE for complete observations (*i.e.* at every generations). Malaspinas *et al.* (2012) approximates the diffusion process by a one-step process. The state space of the process is the population allele frequencies that are denoted by $(z_0,...,z_{H-1})$, where $z_0=0$, $z_{H-1}=1$ and $z_{k-1}<z_k$. The one-step process only allows transitions between two adjacent states (i.e., from

$z_i$ to $z_{i-1}$, or $z_i$ to $z_{i+1}$, hence the infinitesimal generator Q can be constructed as a tridiagonal *HxH* matrix:

$$Q = \begin{pmatrix} 0 & 0 & 0 & & \cdots & & 0 \\ \delta_1 & \eta_1 & \beta_1 & & & & \\ 0 & \ddots & \ddots & \ddots & & & \vdots \\ & & \delta_i & \eta_i & \beta_i & & \\ \vdots & & & \ddots & \ddots & \ddots & 0 \\ & & & & \delta_{H-2} & \eta_{H-2} & \beta_{H-2} \\ 0 & & \cdots & & 0 & 0 & 0 \end{pmatrix}$$

where $\beta_i$ denotes the transition rate from $z_i$ to $z_{i+1}$, $\delta_i$ the transition rate from to $z_i$ to $z_{i-1}$, and $\eta_i$ is the rate of no transition such that $\eta_i=1-(\beta_i+\delta_i)$. The appropriate choice for the parameters $\beta$ and $\delta$ of the matrix $Q$ are given for both the diploid and the haploid Wright-Fisher model in Malaspinas *et al.* (2012) and Foll *et al.* (2014), respectively.

**Simulated datasets for testing**

For real data, it is important to take into account the non-random criteria one used to select sites from the genome for analyses. This so-called ascertainment bias is known to be very important for single time point SNP data (Nielsen and Signorovitch 2003), but has not been studied so far for time-sampled data. One of the reasons is that including realistic ascertainment schemes in likelihood-based methods is a difficult task. The one-step process used in Malaspinas *et al.* (2012) can be adjusted in order to match the way in which ancient DNA data is generally collected, such that the locus considered is polymorphic at the present time. This condition implies that the process can never reach the absorbing states 0 and 1, and one needs to remove the first and last rows and columns of the *Q* matrix. In the current implementation of Malaspinas *et al.* (2012) only this conditional case is available and has been tested for this study. The Bollback *et al.* (2008) method implements an unconditional model (no ascertainment) as well as the particular case where the allele is known to be beneficial and reaches fixation during the sampling period. There is no ascertainment model implemented in Mathieson and McVean (2013) method. One distinct advantage of this simulation-based approach is the ability to easily incorporate different ascertainment schemes into the estimation procedure, as one simply needs to be able to simulate them. We here present three such non-exclusive schemes: (i) observing a minimum allele frequency over the entire trajectory, (ii) observing a minimum allele frequency at the last time point (including fixation like in the Bollback *et al.* (2008) method), and (iii) being polymorphic at the last time point (like in the Malaspinas *et al.* (2012) method). We note in particular that the first case is something that will be present in any data set but is not available in likelihood-based methods. Because the three available likelihood methods implement different ascertainment processes, and these processes lead to more or less informative data, it is not possible to make a direct comparison of their performance. For this reason, we separately compare them with WFABC.

For this comparative study, we generate simulated datasets using the Wright-Fisher model with a range of parameter values for the effective population size ($N_e$) and selection coefficient (*s*). For the diploid Wright-Fisher model, co-dominant time-serial allele frequency data from 1000 replicates for $N_e$=(200, 1000, 5000) and $s \in [0, 0.4]$ are generated. In order to assess the precision and accuracy of these methodologies in estimating two potential empirical cases of small *s* values and large *s* values, the selection coefficients are divided into two sets of *s*=(0, 0.005, 0.01, 0.015, 0.02) and *s*=(0, 0.1, 0.2, 0.3, 0.4). For WFABC, we retained the best 1% of 500'000 simulations based on the Euclidian distance between the observed and simulated *Fsd'* and *Fsi'* statistics and use the mean of the posterior distributions obtained for *s* using a rejection ABC algorithm (Sunnåker *et al.* 2013) as a point estimate. Firstly, we use these simulated datasets to demonstrate the performance of WFABC when different sampling time points and different sample sizes are used. Secondly, we show the influence of the ascertainment procedure using two examples: an unconditional but unrealistic case where all trajectories start with an initial minor allele frequency of 10%, and one ascertained case where a new mutation occurs at the first generation and only the trajectories reaching a frequency of 5% at least in one sampling time point are kept. We use these simulated datasets to compare the performance of WFABC with the method of Mathieson and McVean (2013).

Finally, we focus on the ascertainment case of the allele segregating at the last sampling time point, as this model is the only realistic one allowing us to compare WFABC with a likelihood method. Depending on the strength of selection and the effective population size, mutations reach fixation more or less rapidly. In order to generate data with the condition of being polymorphic at the last sampling time point, the number of generations is adjusted to have a non-zero probability for this condition, allowing us to efficiently simulate such scenarios. A new mutation occurs at the first generation and 100 samples are drawn randomly through binomial sampling, with 12 sampling time points. Using these simulated datasets, comparative studies of WFABC with Malaspinas *et al.* (2012) and Bollback *et al.* (2008) methods are carried out, with the search range and the prior for the selection coefficient set as $s \in [-0.1, 0.1]$ for small values ranging from 0 to 0.02, and $s \in [-0.2, 0.6]$ for large values ranging from 0 to 0.4. Simulated datasets from the haploid Wright-Fisher model are also generated, but only for one set of $N_e = 1000$ in order to validate the performance of the modified haploid version of Malaspinas *et al.* (2012) model described in Foll *et al.* (2014). For Malaspinas *et al.* (2012)*,* we used the quadratic grid option (Gutenkunst *et al.* 2009) for computing the likelihood and the Nelder-Mead simplex algorithm option to find the maximum likelihood (Nelder and Mead 1965).

Both the Bollback *et al.* (2008) and Malaspinas *et al.* (2012) methods estimate parameters based on a single allele frequency trajectory, which thus contains limited information about $N_e$. WFABC utilizes multiple sites in order to estimate $N_e$, which is then used as a prior to estimate selection coefficients. In order to implement an equal comparison, we fix $N_e$ to its true value in all scenarios and evaluate the ability of these approaches to estimate selection coefficients.

In order to demonstrate the advantage of estimating $N_e$ correctly, a final multi-locus scenario is generated where both *s* and $N_e$ values are inferred. For $N_e$ fixed at 1000, 10'000 trajectories are simulated with 500 being under selection with *s* randomly chosen in $[0.05, 0.4]$. We used a search range of $\gamma \in [0, 2000]$ and $N_e \in [50, 2000]$ for the Malaspinas *et al.* (2012) method, and a uniform prior $s \in [-0.2, 0.6]$ for WFABC. For WFABC, all the 10'000 trajectories are used in the first step to obtain a posterior distribution for $N_e$, which is used in the second step to estimate *s* at each locus individually as explained above.

## Results

### Performance of the examined estimation procedures

The performance studies of WFABC are presented in a standard boxplot with the box as the first, second, and the third quartile, and the whiskers as the lowest and highest datum within the 1.5 interquartile range of the lower and upper quartile, respectively. The first boxplot shows the estimated selection coefficients for different numbers of sampling time points as 12, 6 or 2 (Figure S1). As expected, a larger number of sampling time points yields a better estimate of *s* – however, WFABC is able to estimate *s* accurately with as small as 2 sampling time points for moderate values of *s*<0.2. The second boxplots shows the estimated selection coefficients from WFABC for the sample sizes of 1000, 100 and 20 with $N_e$=1000 (Figure S2). The estimation of *s* improves as the number of sample sizes increases as expected.

For the comparative studies of WFABC with the Mathieson and McVean (2013) method, the unconditional case with an initial minor allele frequency of 10% is shown in boxplots for the small *s* values (Figure S3) and the large *s* values (Figure S4). For the small *s* values, the Mathieson and McVean (2013) method performs better than WFABC. However, for larger values of *s*, the Mathieson and McVean (2013) method shows an increasing trend of underestimation, whereas WFABC remains unbiased (Figure S4). For the conditional case of ascertaining the simulated datasets with a minimum frequency of 5% at one sampling time point, the performance of WFABC is noticeably better for both the small and large *s* values (Figure 1, Figure 2). Figure 1 shows that the Mathieson and McVean (2013) method constantly overestimates small *s* values. Figure 2 indicates that this biased is compensated by the underestimation shown above for large *s* values.

For the conditional case studies of the allele segregating at the last sampling time point, we obtain 6 sets of results with the varying parameters for the diploid model and 2 sets of results for the haploid model from the three approaches. Results are given for $N_e$ = 1000 for small *s* values (Figure 3), and large *s* values (Figure 4). Additionally, tables with the calculations of the root mean square error (RMSE) and bias for each set of results are shown for the small *s* values and the large *s* values in Table 1 and Table 2, respectively. Note that MSE is defined as the sum of the variance and the squared bias of the estimator, and therefore incorporates information from both precision (variance) and accuracy (bias).

Comparing the three approaches for small *s* values (Figure 3), WFABC and the Malaspinas *et al.* (2012) approach produce good estimates of *s*, whereas the Bollback *et al.* (2008) method fails to infer different *s* values. From the boxplot, both the median and the mean of WFABC and the Malaspinas *et al.* (2012) approach are close to the true *s* value. However, WFABC appears to contain a longer tail of under-estimated *s* values, and the interquartile range boxes are wider. Table 1 provides more quantitative comparisons of their performance. The RMSE values reveal that for all the cases of $N_e$=1000 and small *s*, the Malaspinas *et al.* (2012) approach is generating more precise estimates. On the other hand, WFABC produces estimates of less bias for all small *s* values in this set.

Comparing the three approaches for large *s* values (Figure 4), WFABC and the Malaspinas *et al.* (2012) approach produce reasonable estimates of *s*, whereas the Bollback *et al.* (2008) method again fails to detect any difference in *s* values. For *s* values larger than 0.1, the performance of WFABC is significantly better than the Malaspinas *et al.* (2012) approach both in accuracy and precision (Table 2) – producing estimates with 10-fold less bias and 2-fold less error than the Malaspinas *et al.* (2012) approach. This gap in performance increases from *s* = 0.2 to *s* =0.4, thus this trend may be extrapolated to higher *s* values.

Notably, the Bollback *et al.* (2008) method consistently estimates *s* = 0 for all examined datasets. This poor performance has been evaluated for various conditions including changing grid sizes, conditioning on fixation, and varying sampling time points – with no perceptible difference in results. In order to verify our usage, the method was tested with the exact parameters utilized in the initial paper for the example of bacteriophage MS2 (Bollback *et al.* 2008), resulting in the successful replication of their results (Figure S5, Figure S6). Notably, the performance of the statistic depends on the choice of the search range for $\gamma$, due to the presence of local peaks in the likelihood function. When the interval is chosen to be narrow and centered around the true *s* value of 0.4, the estimated *s* is correctly given owing to the local maximum (replicating their result). However, when the full likelihood surface is examined, the global maximum is present near 0 as observed in all simulated test replicates. For this reason, the Bollback *et al.* (2008) method is excluded from the further analyses of performance, as well as from the illustrative data application.

As shown, our comparison studies suggest that WFABC and Malaspinas *et al.* (2012) approaches perform almost equally well for estimating selection coefficients for both small and large values. While WFABC slightly overestimates the selection coefficient in the cases of small *s* and small $N_e$ (Figure S7, Table S1), it is notable that the Malaspinas *et al.* (2012) performs particularly well when *s* is in the range of 0.01 and 0.02, as observed by Malaspinas *et al.* (2012). In contrast, WFABC exhibits less bias when *s* values are large (i.e., greater than 0.1) for small $N_e$ (Figure S8, Table S2). Thus, we conclude that the two methods estimate the selection coefficient to a high accuracy and precision for $N_e \in [200,1000]$ and for the condition of the allele segregating at the last sampling time point. In general, although the difference in performance both in precision and accuracy is minor, the Malaspinas *et al.* (2012) approach appears to give superior results for small *s* values, whereas WFABC for large *s* values.

However, the Malaspinas *et al.* (2012) approach encounters a limitation in computation efficiency for the set of $N_e$=5000 (Figures S9, S10; Tables S3, S4). For the large *s* values, the computation time was too lengthy to complete the 1000 replicates (see below), thus only the results from WFABC are shown (Figure S10). Compared to the small $N_e$ values, the estimation of large *s* is becoming more accurate and precise as $N_e$ gets large (Tables 2, S2, S4). This trend is the same for small *s* values (Tables 1, S1, S3), although the bias appears to switch from over-estimation to under-estimation as $N_e$ increases. For the Malaspinas *et al.* (2012) approach, the accuracy of inferring small *s* values improves as $N_e$ increases, although the precision remains similar (Tables 1, S1, S3). However, for $N_e$ =5000 and *s*>0.01, the 1000 replicates were not complete due to the lengthy computational time, thus the RMSE and bias values are not available (Table S3).

For the haploid model, WFABC shows superiority in both accuracy and precision for inferring any *s* values compared with the Malaspinas *et al.* (2012) approach (Figures S11 and S12, Tables S5 andS6).

Finally, the multi-locus scenario demonstrates the great benefit provided by the ability of WFABC to use the information shared by all loci to estimate $N_e$. The RMSE of the selection coefficients calculated over the 500 trajectories under selection is less than half that obtained using the Malaspinas *et al.* (2012) approach (0.049 vs. 0.10, see Figure 5).

In summary, WFABC is superior for diploid cases when both *s* and $N_e$ values are large (i.e., *γ=2$N_e$s* is large), for any haploid cases, and when multiple loci are available, whereas the Malaspinas *et al.* (2012) approach is suitable for cases when *γ* values are small.

**Comparison of computational efficiency**

Apart from accuracy and precision, an important difference in performance between these methods is the computational efficiency. A major advantage of WFABC is the computational speed, which, for example, allows for an evaluation of all observed sites in the genome in order to identify putatively selected outliers (Foll *et al.* 2014), and to estimate the full distribution of fitness effects of segregating mutations. For the Malaspinas *et al.* (2012) approach, the computational time becomes heavy when $\gamma$ is larger than 200 and is no longer feasible when $\gamma$ is approaching 1000, whereas WFABC has no restriction on the sizes of $N_e$ and $s$. For the Mathieson and McVean (2013) approach, the CPU time of estimating only the selection coefficient of each site is around 2 seconds regardless of the size of $N_e$, but still is slower than WFABC. Therefore, we suggest that the likelihood-based approach is preferable in cases where both the candidate mutation and effective population sizes are known *a priori*, whereas WFABC is preferable in the absence of this information. The average CPU time spent for each replicate for the diploid model is shown in Table 3. We note that when the Malaspinas *et al.* (2012) method is also used to estimate $N_e$, the difference in CPU time between the two methods is even greater.

**Data Application**

We applied WFABC to a time-serial data set of the *medionigra* morph in a population of *Panaxia dominula* (scarlet tiger moth) at Cothill Fen near Oxford. This colony was first studied by Fisher and Ford (1947), and further observations have been collected almost every year until at least 1999 (Cook and Jones 1996; Jones 2000). The moth *P. dominula* has a one year generation time and lives near the Oxford district in isolated colonies. The typical phenotype has a black fore wing with white spots and a scarlet hind wing with black patterns (see Fisher and Ford 1947). The *medionigra* allele produces the *medionigra* phenotype when heterozygous, and the *bimacula* phenotype when homozygous, changing the pigment and patterns on the wings to an increasing degree, and is almost never observed (Sheppard and Cook 1962). Using our notation above, we denote by *A* the *medionigra* allele and the fitness of the three genotypes are given by $w_{AA}=1+s$ (*bimacula*), $w_{Aa}=1+sh$ (*medionigra*) and $w_{aa}=1$.

The respective role of drift and natural selection to explain the rapid decline of the *medionigra* allele frequency after 1940 (Figure 6) was the subject of a strong debate between Fisher and Wright (Fisher and Ford 1947; Wright 1948), with Wright arguing that the observed pattern until 1946 could be explained by drift alone with an effective population size of $N_e$=150 (Wright 1948). The same data containing further observations has been re-analyzed several times (Cook and Jones 1996; O'Hara 2005; Mathieson and McVean 2013) with most studies concluding that the *medionigra* allele is negatively selected with s=-0.14 (Cook and Jones 1996) or s=-0.11 (Mathieson and McVean 2013) based on a co-dominant model (*h=1/2*). Recently, Mathieson and McVean (2013) found that a fully recessive *medionigra* allele (*h=0*) fits with a higher likelihood compared to *h=1/2* but with a much larger selection coefficient *s≈-1*. In particular, this recessive lethal model explains better the persistence of the *medionigra* allele at a low frequency for so many generations (Mathieson and McVean 2013). However this large value of *s* is outside the range for which their approximations are valid and this hypothesis could not be formally tested.

Our ABC approach based on simulations can deal with the full range of *s* values and we further extended it here to co-estimate the degree of dominance *h*. We followed the intuitive idea of Mathieson and McVean (2013) that the number of generations during which the allele persists at low frequency is informative for the degree of dominance *h.* More formally, we added two summary statistics in our ABC procedure, *tl*, defined as the number of generations where the allele frequency is below 5% and not lost; and *th* defined as the number of generations where the allele frequency is above 95% and not fixed. For the moth data, we have *tl=54* and *th=0* (see Figure 6). Using the simulated distributions, *Fsd'* and *Fsi'* are both normalized by the largest standard deviation *max(sd(Fsd'),sd(Fsi'))*, as well as *tl* and *th* by *max(sd(tl),sd(th))*. We followed Mathieson and McVean (2013) and ran our ABC method using a fixed population size of *2$N_e$=1000* and we plot the corresponding joint posterior distribution for *s* and *h* in Figure 7A. The mode of the joint posterior distribution is at *s=-1* and *h=0.043*, supporting the idea of a lethal *bimacula* phenotype (*$w_{AA}$=0*) with a deleterious *medionigra* phenotype (*$w_{Aa}$=0.96*) consistent with previous observations (Sheppard and Cook 1962). The shape of the joint posterior distribution (Figure 7A) shows that the *medionigra* allele is either very strongly selected against and almost completely recessive (*h≈0*), or co-dominant with a weaker selection coefficient (*s≈-0.2*). Even if the density is larger in the recessive lethal region of the parameter space (bottom left corner in Figure 7A), the two-dimensional 90% highest posterior region includes the alternative co-dominant hypothesis. As it has been argued that the effective population size could be of the order of a few hundred (Wright 1948; O'Hara 2005), we also ran the analysis with *2$N_e$=100*. The joint posterior distribution for *s* and *h* in Figure 7B shows a similar pattern with a mode at *s=-1* and *h=0*. However the surface is flatter and the two-dimensional 90% highest posterior region now includes *s=0,* confirming Wright's view that a small enough population size could explain the observed pattern with genetic drift alone (Wright 1948; Mathieson and McVean 2013). We finally ran the analysis using a uniform prior for $2N_e$ between 100 and 10000 to take into account its uncertainty in the estimate. In this case the joint posterior gives a stronger support for a lethal *bimacula* phenotype with a deleterious *medionigra* phenotype as compared to $2N_e=1000$ (Figure S13).

**Discussion**

Maximum likelihood estimators (MLE) have the advantage of being consistent and efficient, but the computational method used to find the maximum likelihood can be critical in complex models. For instance, even though the underlying model of the Bollback *et al.* (2008) approach is similar to the Malaspinas *et al.* (2012) approach and the Mathieson and McVean (2013) approach, the difference in performance appears to be coming from the difference in the implementation of computational methods. On the other hand, ABC-based methods reduce datasets into summary statistics, and thus the performance of these methods is dependent on the chosen statistics. The difference in performance demonstrated through this comparative study between our newly proposed ABC-based method and the likelihood-based methods is very small in most cases. Therefore, we can hypothesize that the two summary statistics - *Fsd'* and *Fsi'* - are close to being statistically sufficient.

A disadvantage of the likelihood methods arises from the limitations imposed by the assumptions made for diffusion approximation. In order to approximate the Wright-Fisher model with the diffusion process, one makes the assumption that the Markov process is continuous in state space and time. This assumption requires $s$ to be small and $N_e$ to be large (Durrett, 2008). Thus, likelihood methods based on diffusion approximation are inevitably limited to the cases of large effective population sizes and small selection coefficients, and may explain the biases observed at large *s* values in this study. Additionally the computational efficiency limits the value of *γ* to be under 1000 for the Malaspinas *et al.* (2012) approach, even if *s* is small, $N_e$ is again limited to be under 5000. Thus, despite having good performance, the likelihood methods are limited to cases of small *s* and intermediate $N_e$ values.

Although WFABC gives slightly less precise results in some cases of small *s*, the two major advantages of this implementation comes from its ability to consider complex ascertainment cases and to accurately estimate $N_e$ from multi-locus genome wide data. The Mathieson and McVean (2013) method is computationally efficient but is based on the unrealistic assumptions that $N_e$ is known and that the data is completely unascertained. Using a point estimate of $N_e$ obtained from the first step of WFABC in the Mathieson and McVean (2013) method would ignore the uncertainty on this parameter. Malaspinas *et al.* (2012) only handles the conditional cases of polymorphism at the last sampling time point and estimates $N_e$ separately at each site. WFABC is computational efficient and enables the estimation of a genome-wide average $N_e$ from time-sampled data, after which per-site selection coefficients may be estimated efficiently and accurately. There currently exist no such likelihood-based methods to estimate $N_e$ accurately from time-sampled genomic data, since the information is only based on a single allele trajectory. Our simulated multi-locus scenario shows that estimating $N_e$ accurately leads to a higher precision in the estimates of selection coefficients *s*, leading overall to a smaller RMSE as compared to the Malaspinas *et al.* (2012) method. As the effective population size is an important population genetic parameter that is often unknown in both ancient genomics and experimental evolution, WFABC provides a practical and flexible platform to be utilized in any time-serial data for efficiently inferring a wide range of $N_e$ and *s* values to high accuracy. Finally, we note that the ABC method also has the advantage of providing a posterior distribution rather than simply a point estimate, allowing for easy-to-build credibility intervals.

The application of WFABC to the *P. dominula* data confirms that a nearly fully recessive lethal model for the *medionigra* allele is the best explanation for the observed pattern as hypothesized by Mathieson and McVean (2013). We note that in this case, once the allele frequency is low enough such that heterozygotes almost never occur, it behaves like a neutral model. We used this feature to estimate $N_e$ using *Fs'* (Jorde and Ryman 2007) by considering only time-points after 1950, when the *medionigra* allele first reaches a frequency below $\sqrt{0.001} \approx 0.032$ and we obtained *$2N_e$=927*, which is consistent with previous estimates (Fisher and Ford 1947; Cook and Jones 1996; O'Hara 2005). This application also demonstrates that WFABC is very flexible, as it can also be used to co-estimate *h* and accommodate very large selection coefficients (such as *s=-1*, as in our application here). In order to confirm the validity of our approach in this case, we simulated 1000 datasets mimicking the *P. dominula* data (same number of generations, time points, sample sizes and initial allele frequency). We fixed *s=-1* and *h=0.05* and let *$2N_e$* vary uniformly between 100 and 10000 for each simulation, and estimated *s* and *h* using our ABC method (Figure 8). Both parameter estimates are unbiased and while distinguishing lethality (*s=-1*) from very strong negative selection (*s<-0.5*) seems to be difficult, *h* is estimated with a very small variance.

Finally, it should be noted that all the methods described and utilized in this study assume that the loci are in linkage equilibrium and take no demographic history into account except the Mathieson and McVean (2013) method which integrates structured populations with a lattice model. Owing to the generally low number of generations in time series data, one expects only little recombination to occur and long stretched of DNA may be under linkage disequilibrium. Foll *et al.* (2014) demonstrated that the model is robust to fluctuating population sizes, but this may not hold for all demographic scenarios. It is an important future challenge to expand these methods further for inferring demographic parameters from time-serial data. WFABC has great potential in achieving this challenging task, as ABC-based estimators lend themselves more readily to the incorporation of complex demographic models compared to likelihood-based methods.

## Acknowledgements

The computations were performed at the Vital-IT (http://www.vital-it.ch) Center for high-performance computing of the SIB Swiss Institute of Bioinformatics. This work was funded by grants from the Swiss National Science Foundation and a European Research Council (ERC) starting grant to JDJ.

## Data Accessibility

This study is primarily based on simulated data created using the WFABC software available from the "software" page at http://jensenlab.epfl.ch/. The *P. dominula* moth data set has been taken from (Cook and Jones 1996; Jones 2000) and is also provided in the WFABC package.

## Author Contributions

M.F. and J.D.J conceived the idea. M.F. developed the WFABC software and H.S. extended the Malaspinas et al. (2012) software for haploids. M.F. performed the simulations and the WFABC analyses. H.S. performed the Mathieson and McVean (2013), Bollback *et al.* (2008) and Malaspinas *et al.* (2012) analyses. All authors contributed to writing the manuscript.

## Tables

**Table 1.** RMSE and bias for the small $s$ and $N_e$=1000 scenario for the Wright-Fisher diploid model

|      |                         | s=0     | s=0.005 | s=0.01  | s=0.015 | s=0.02  |
|------|-------------------------|---------|---------|---------|---------|---------|
| Bias | WFABC                   | -0.0035 | -0.0046 | -0.0032 | -0.0014 | -0.0015 |
|      | Malaspinas *et al.* (2012) | -0.0044 | -0.0050 | -0.0039 | -0.0026 | -0.0022 |
| RMSE | WFABC                   | 0.017   | 0.018   | 0.017   | 0.016   | 0.018   |
|      | Malaspinas *et al.* (2012) | 0.013   | 0.014   | 0.012   | 0.011   | 0.011   |

**Table 2.** RMSE and bias for the big $s$ and $N_e$=1000 scenario for the Wright-Fisher diploid model

|      |                         | s=0     | s=0.1   | s=0.2   | s=0.3    | s=0.4    |
|------|-------------------------|---------|---------|---------|----------|----------|
| Bias | WFABC                   | 0.017   | -0.017  | 0.0064  | -0.0044  | -0.0083  |
|      | Malaspinas *et al.* (2012) | -0.0024 | -0.018  | -0.025  | -0.056   | -0.091   |
| RMSE | WFABC                   | 0.059   | 0.060   | 0.050   | 0.046    | 0.059    |
|      | Malaspinas *et al.* (2012) | 0.047   | 0.046   | 0.048   | 0.084    | 0.13     |

**Table 3.** Average CPU time consumed in seconds to run one simulation replicate.

|                         | $N_e$=200 | | $N_e$=1000 | | $N_e$=5000 | |
|-------------------------|-----------|---------|-----------|---------|-----------|---------|
|                         | Small s   | Large s | Small s   | Large s | Small s   | Large s |
| WFABC                   | 5         | 3       | 15        | 6       | 300       | 70      |
| Malaspinas *et al.* (2012) | 380       | 250     | 350       | 18000   | 80000     | -       |

**Figure 1.** Boxplot for the estimated small selection coefficients from each simulated replicate of the Wright-Fisher diploid model, with $N_e$=1000, 90 generations, and the minimum frequency of 5% at one sampling time point. The gray rectangles correspond to WFABC and the white rectangles to the Mathieson and McVean (2013) method. The red circles are the true values for *s*, and the blue triangles are the mean of the estimated *s* values.

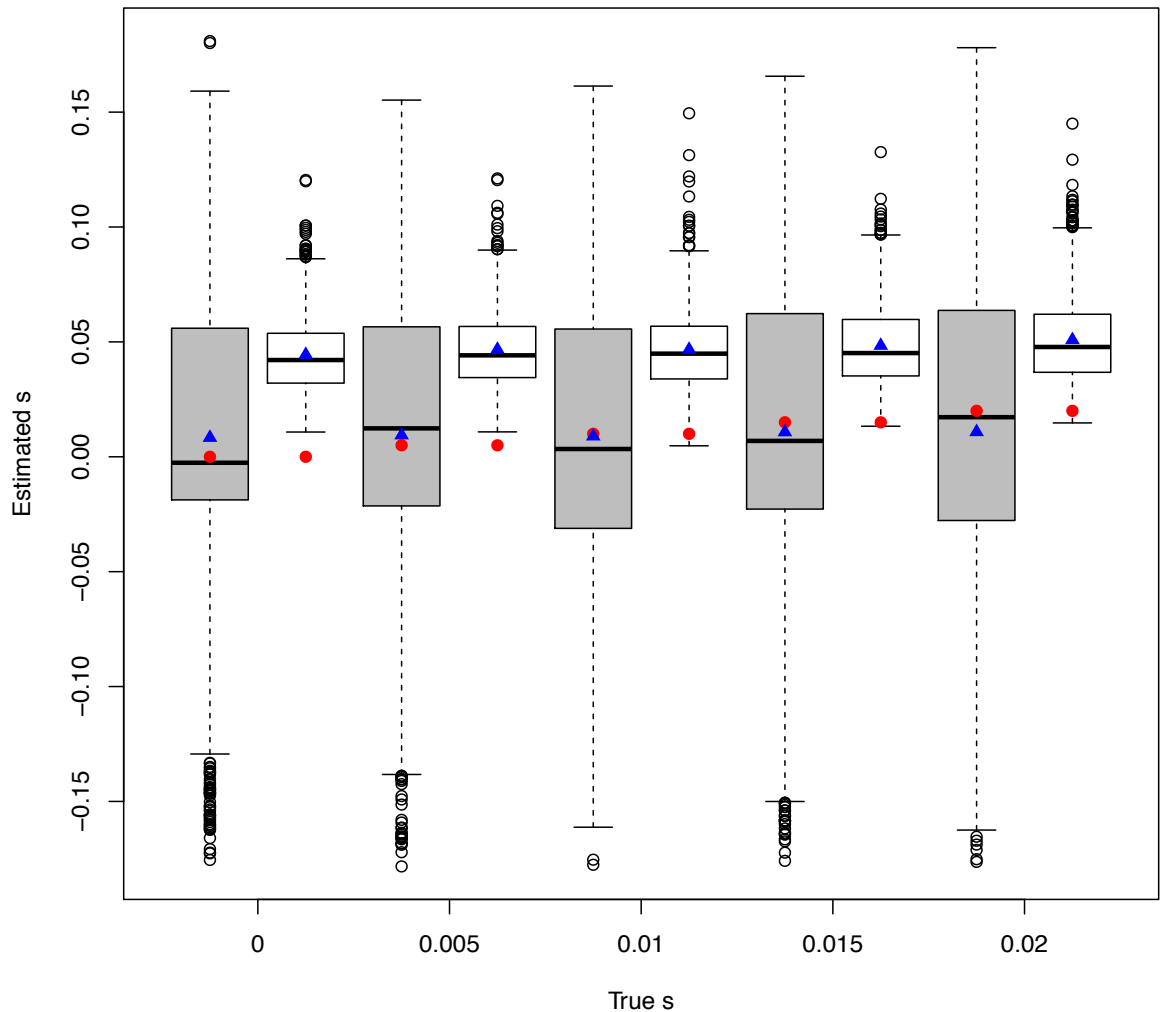

**Figure 2.** Boxplot for the estimated large selection coefficients from each simulated replicate of the Wright-Fisher diploid model, with $N_e$=1000, 90 generations and the minimum frequency of 5% at one sampling time point. The gray rectangles correspond to WFABC and the white rectangles to Mathieson and McVean (2013) method. The red circles are the true values for *s*, and the blue triangles are the mean of the estimated *s* values.

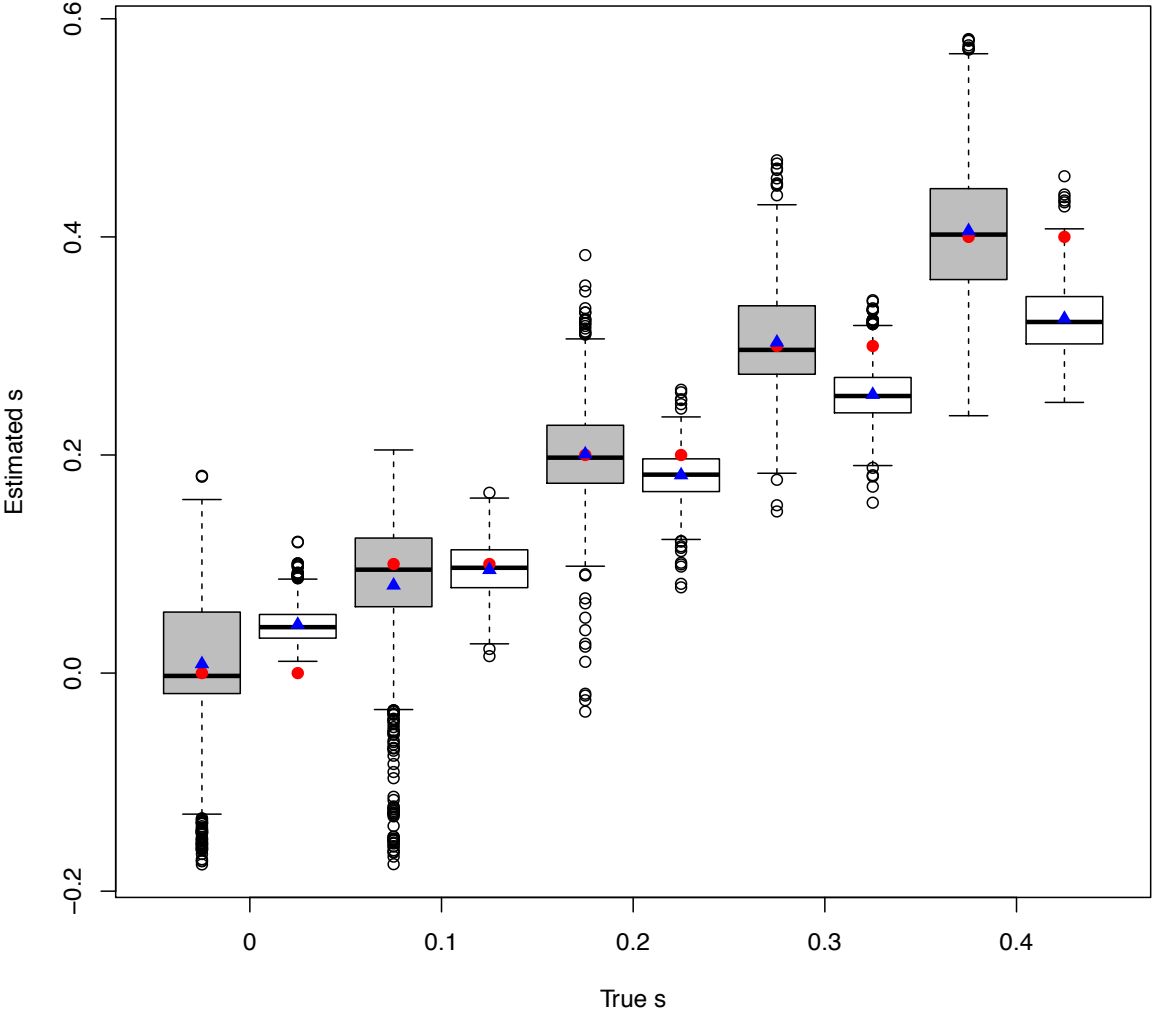

**Figure 3.** Boxplot for the estimated small selection coefficients from each simulated replicate of the Wright-Fisher diploid model, for $N_e$=1000 and 500 generations. The gray rectangles correspond to WFABC, the white rectangles to the Malaspinas *et al.* (2012) method, and the gold rectangles to the Bollback *et al.* (2008) method. The red circles are the true values of *s*, and the blue triangles are the mean of the estimated *s* values.

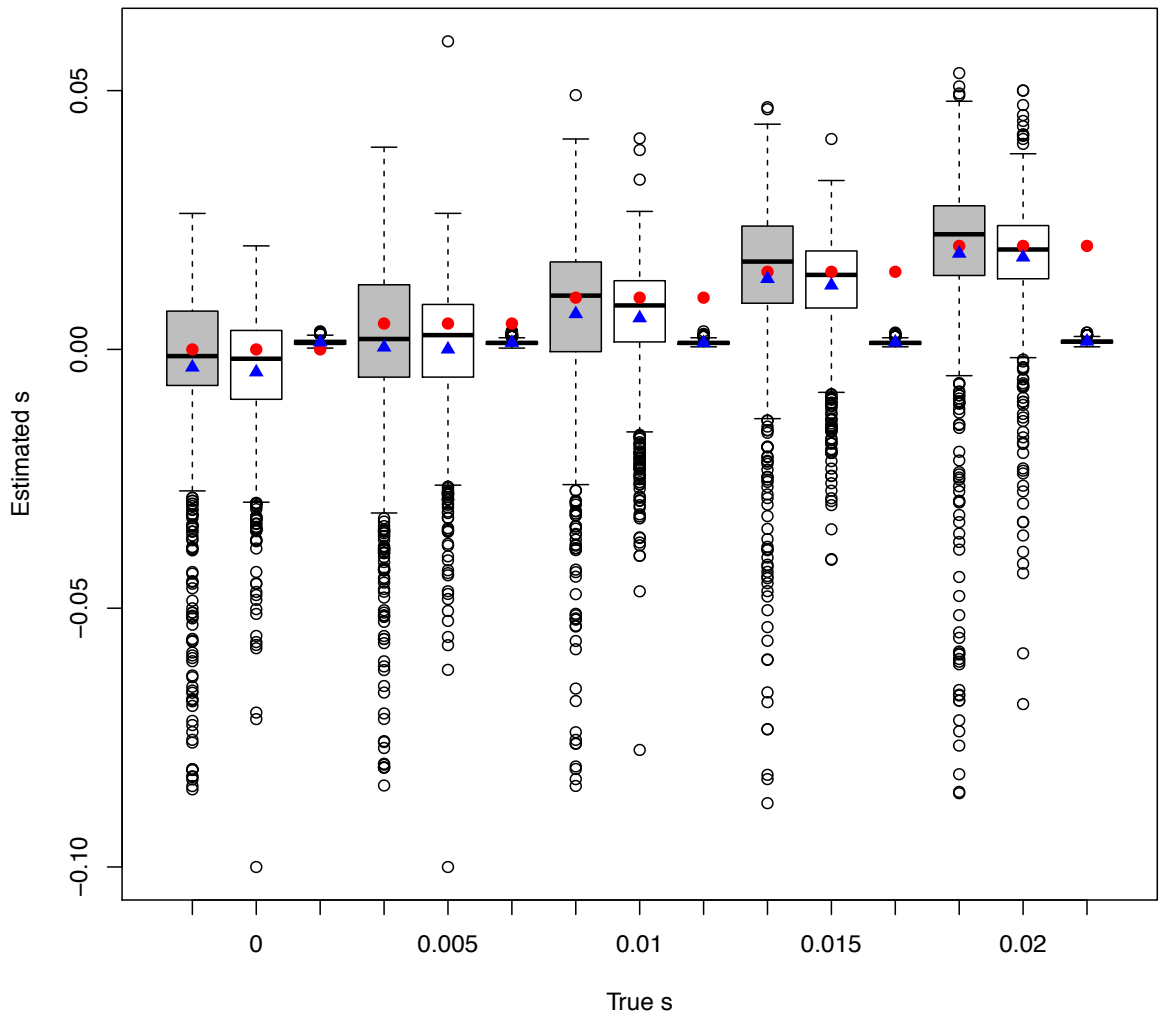

**Figure 4.** Boxplot for the estimated large selection coefficients from each simulation replicate of the Wright-Fisher diploid model, for $N_e$=1000 and 90 generations. The gray rectangles correspond to WFABC, the white rectangles to the Malaspinas *et al.* (2012) method, and the gold rectangles to the Bollback *et al.* (2008) method. The red circles are the true values for *s*, and the blue triangles are the mean of the estimated *s* values.

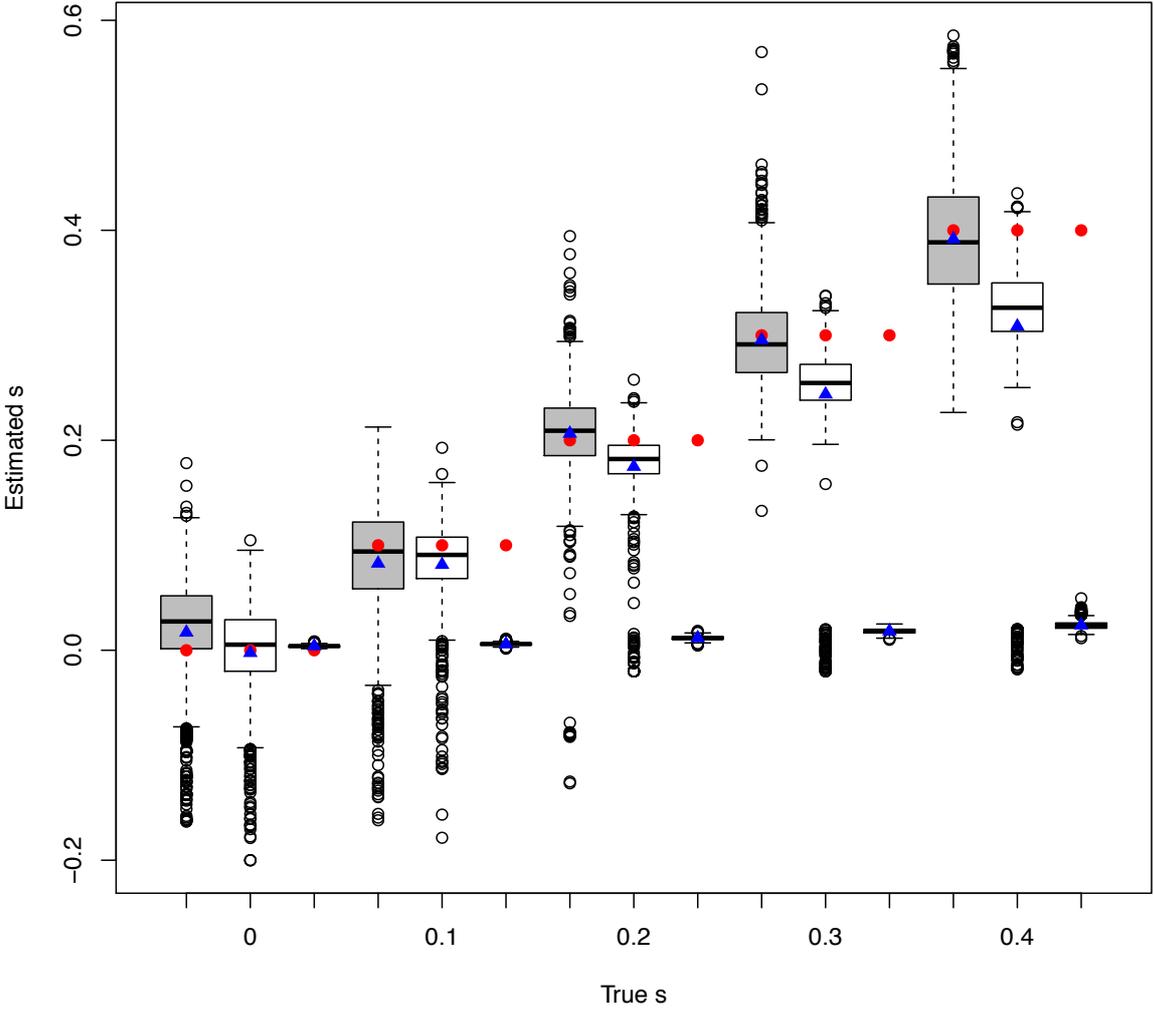

**Figure 5.** Correlation between the true and the estimated values of the selection coefficients *s* for the multi-locus scenario using WFABC (A) and the Malaspinas *et al.* (2012) method (B). The red line represents the identity (*y=x*) line.

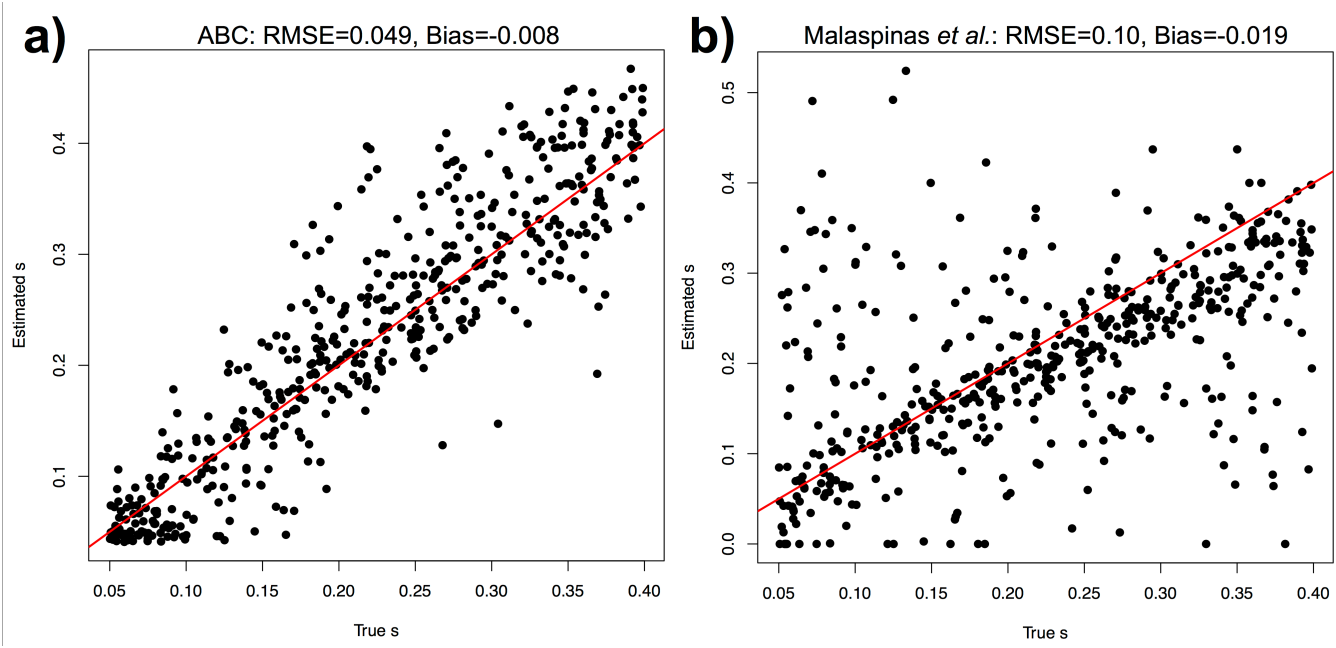

**Figure 6.** Frequency of the *medionigra* allele in the Cothill *P. dominula* moth population (Cook and Jones 1996; Jones 2000).

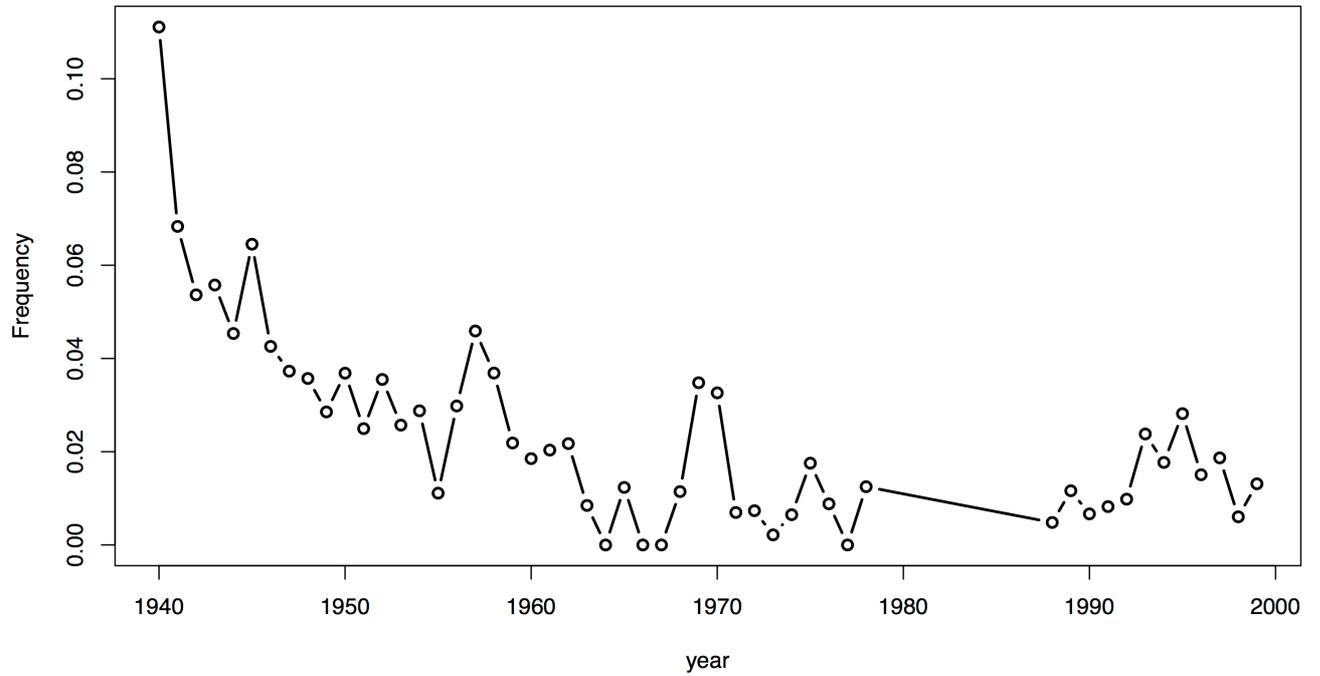

**Figure 7.** Two-dimensional joint posterior distribution for *s* and *h* for the moth *P. dominula* data using *2N$_e$=1000* (a) and *2N$_e$=100* (b). The grey lines delimit two-dimensional $\alpha$ highest posterior density regions for $\alpha$ =0.9 (largest region), 0.8, 0.7, 0.6, 0.5, 0.4, 0.3, 0.2 and 0.1 (smallest region).

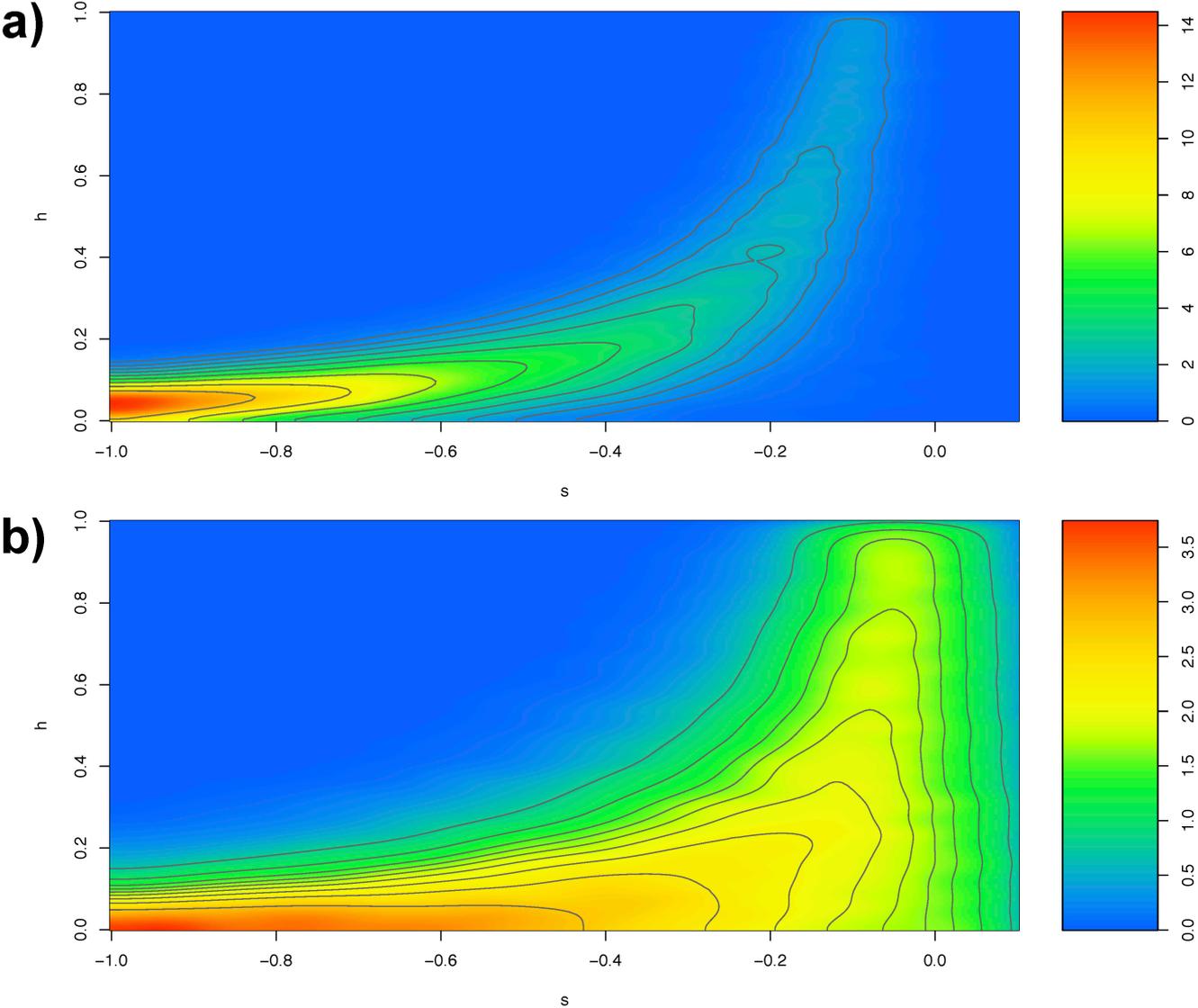

**Figure 8.** Boxplot for the estimated selection coefficient (*s*, panel a) and dominance ratio (*h*, panel b) from 1000 simulated datasets mimicking the *P. dominula* data. The red circles indicate the true simulated values (*s=-1; h=0.05*), and the blue triangles the mean of the estimated values.

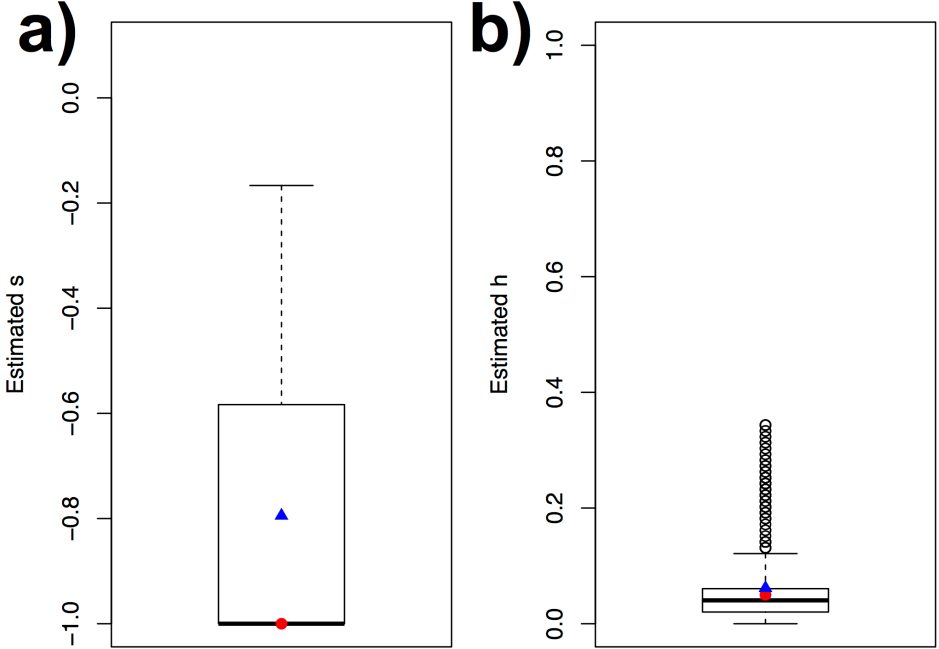

**Figure S1.** Boxplot for the estimated selection coefficient from WFABC for 3 different sampling time points. Each simulation replicate is from the Wright-Fisher diploid model with $N_e=1000$ simulated for 90 generations. The dark gray rectangles correspond to 12 sampling time points, the light gray rectangles to 6 sampling time points, and the white rectangles to 2 sampling time points. The red circles are the true values for *s*, and the blue triangles are the mean of the estimated *s* values.

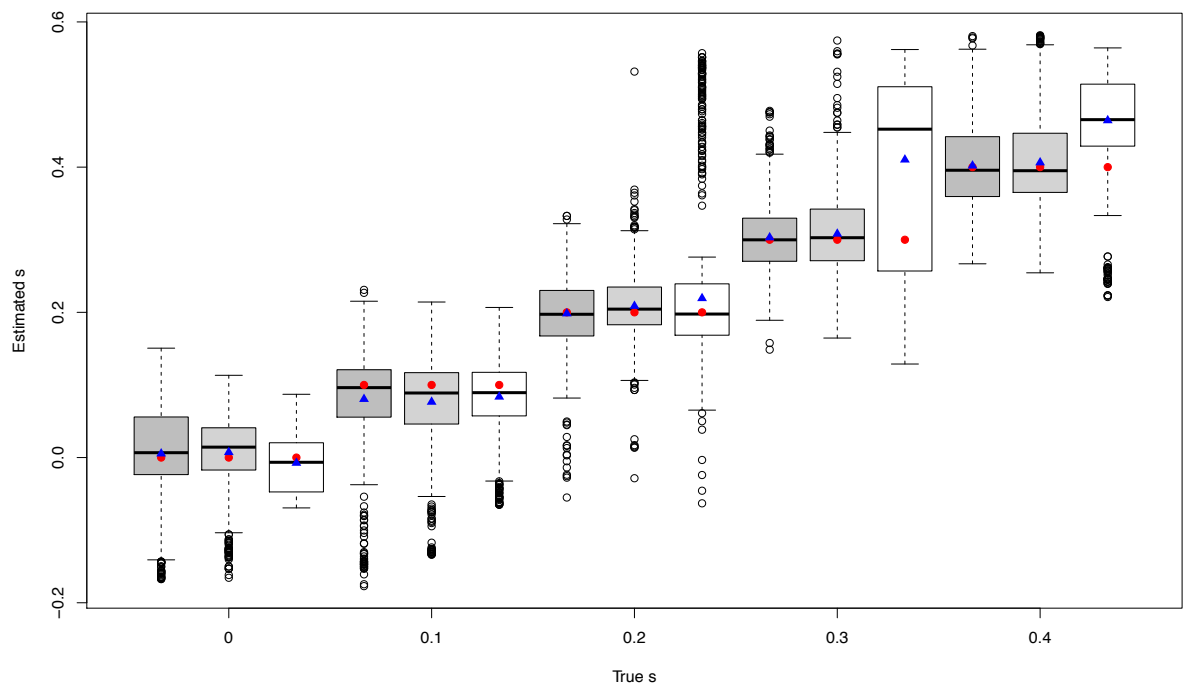

**Figure S2.** Boxplot for the estimated selection coefficient from WFABC for 3 different sample sizes. Each simulation replicate is from the Wright-Fisher diploid model with $N_e$=1000 simulated for 90 generations. The dark gray rectangles correspond to 1000 samples, the light gray rectangles to 100 samples, and the white rectangles to 20 samples. The red circles are the true values for *s*, and the blue triangles are the mean of the estimated *s* values.

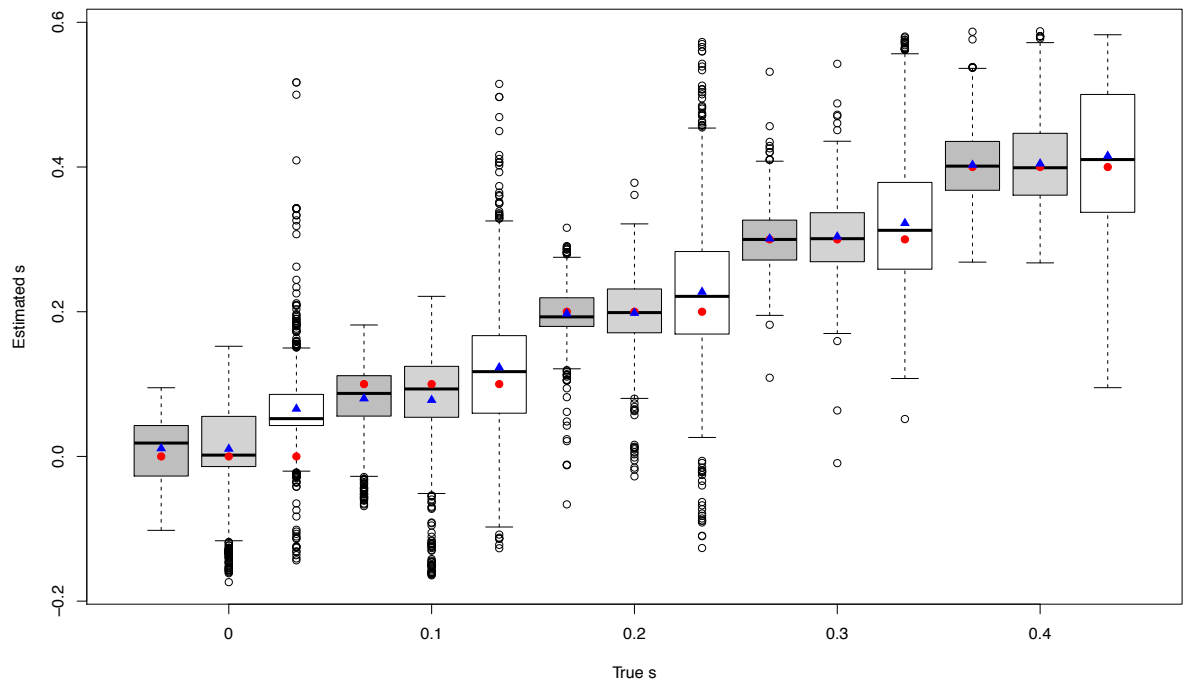

**Figure S3.** Boxplot for the estimated small *s* from each simulation replicate of the Wright-Fisher diploid model with $N_e$=1000 simulated for 90 generations and the initial minor allele frequency at 10%. The gray rectangles correspond to WFABC and the white rectangles to the Mathieson *et al.* (2013) method. The red circles are the true values for *s*, and the blue triangles are the mean of the estimated *s* values.

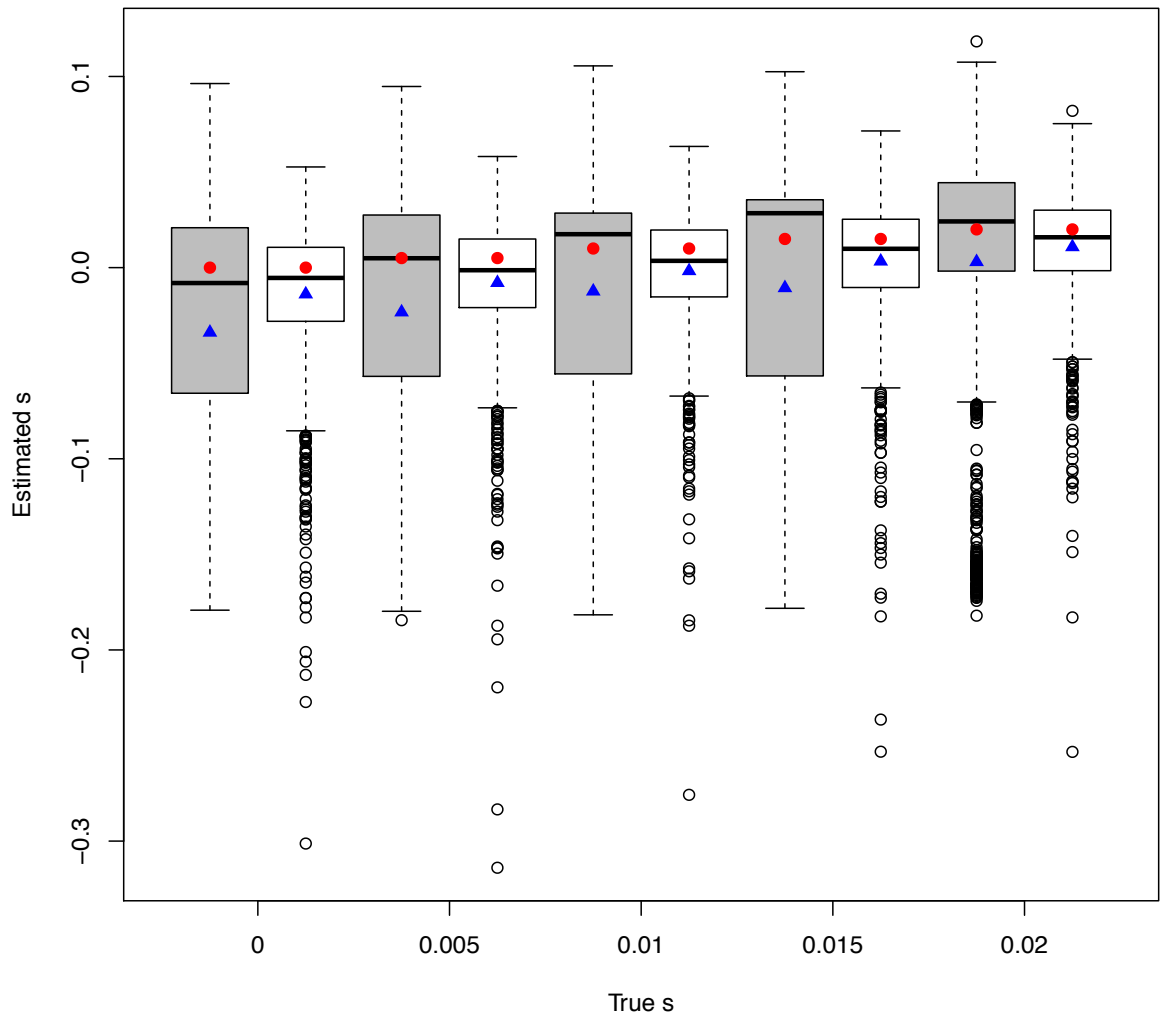

**Figure S4.** Boxplot for the estimated large *s* from each simulation replicate of the Wright-Fisher diploid model with $N_e=1000$ simulated for 90 generations and the initial minor allele frequency at 10%. The gray rectangles correspond to WFABC and the white rectangles to the Mathieson *et al.* (2013) method. The red circles are the true values for *s*, and the blue triangles are the mean of the estimated *s* values.

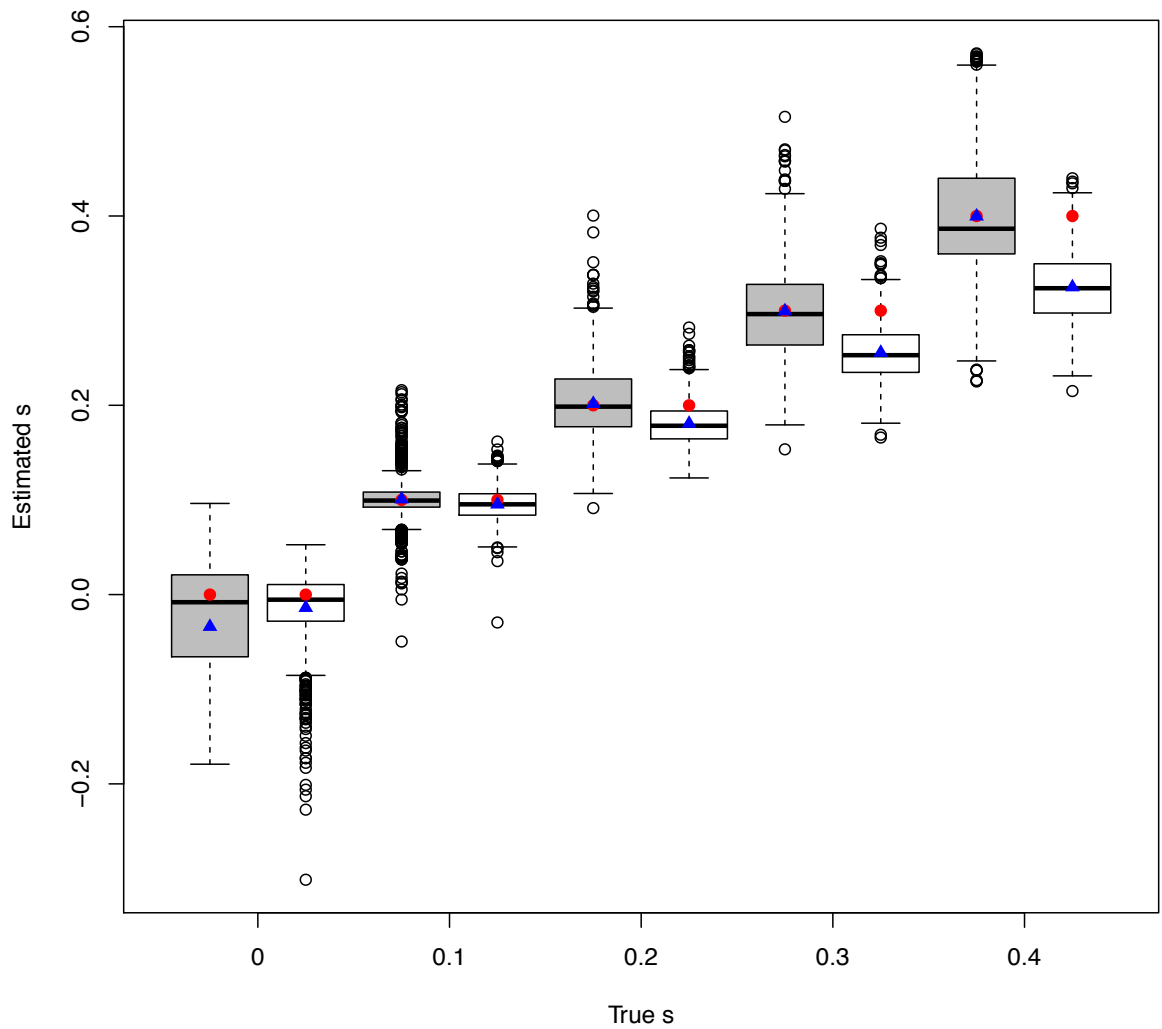

**Figure S5.** Log likelihood of the estimated selection coefficient for the Bollback *et al.* (2008) method with a small search interval for γ. The input trajectory from the Wright-Fisher diploid model with $N_e$=1e+07 and *s*=0.4 simulated for 120 generations, and 50 sampled chromosomes.

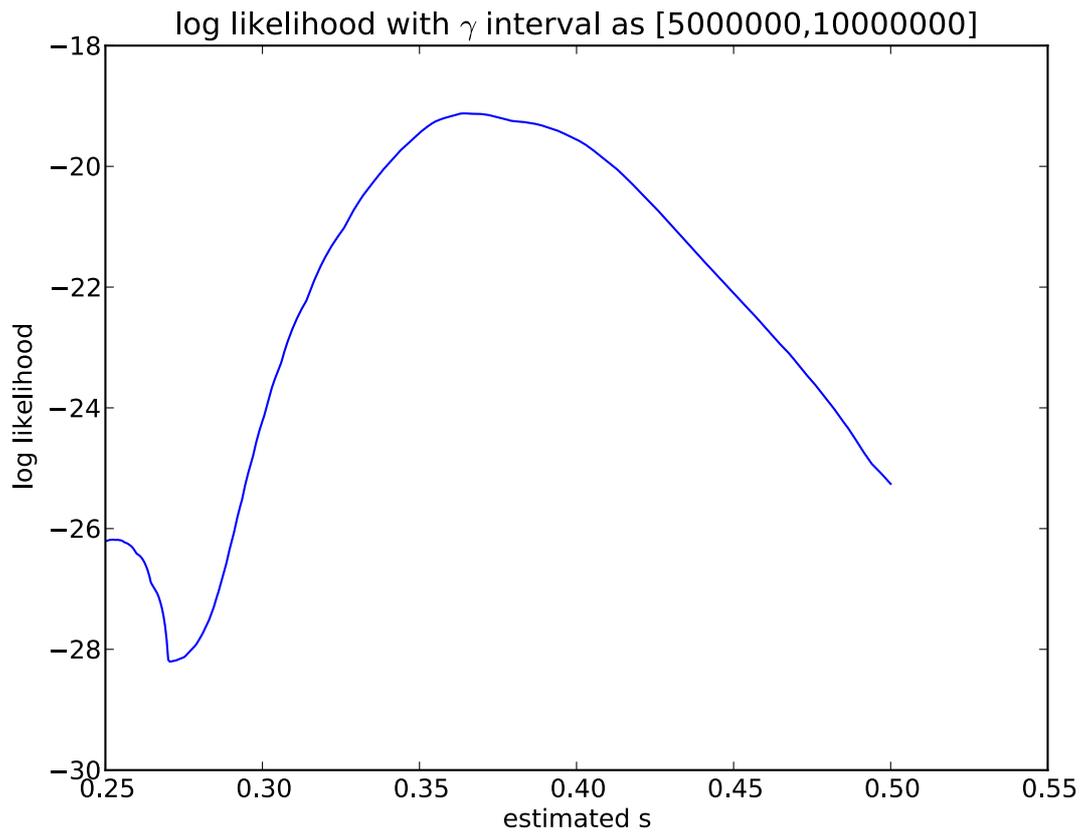

**Figure S6.** Log likelihood of the estimated selection coefficient for the Bollback *et al.* (2008) method with a large search interval for γ. The input trajectory from the Wright-Fisher diploid model with $N_e$=1e+07 and *s*=0.4 simulated for 120 generations, and 50 sampled chromosomes.

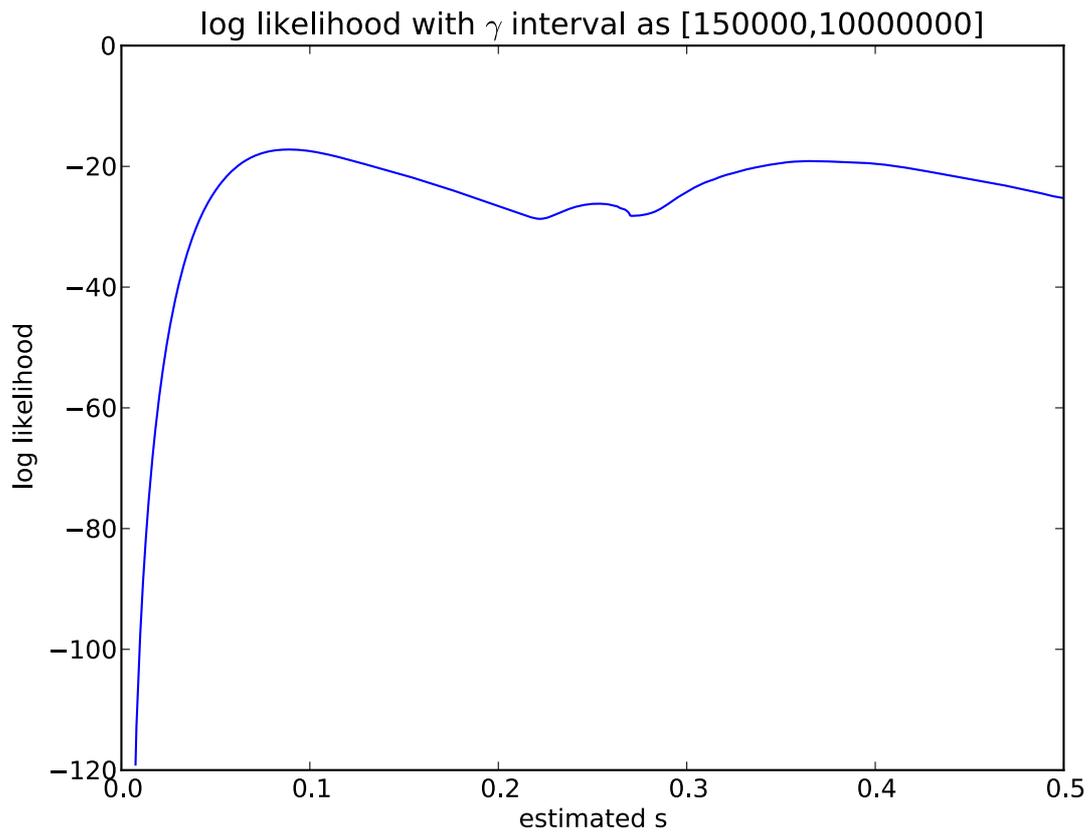

**Figure S7.** Boxplot for the estimated small selection coefficient from each simulation replicate of the Wright-Fisher diploid model with $N_e$=200 simulated for 300 generations. The gray rectangles correspond to WFABC, and the white rectangles to the Malaspinas *et al.* (2012) method. The red circles are the true values for *s*, and the blue triangles are the mean of the estimated *s* values.

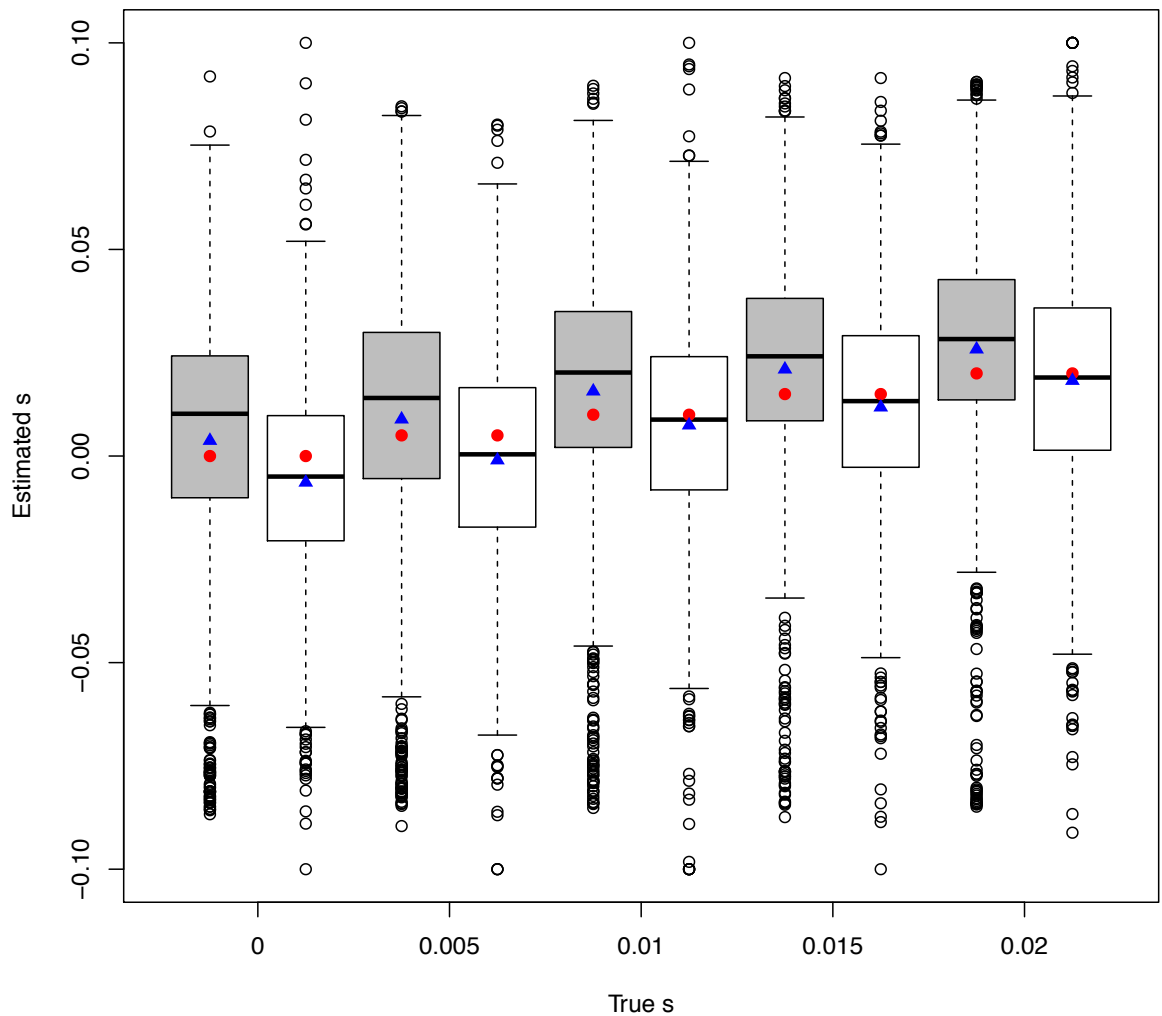

**Figure S8.** Boxplot for the estimated large selection coefficient from each simulation replicate of the Wright-Fisher diploid model with $N_e$=200 simulated for 80 generations. The gray rectangles correspond to WFABC, and the white rectangles to the Malaspinas *et al.* (2012) method. The red circles are the true values for *s*, and the blue triangles are the mean of the estimated *s* values.

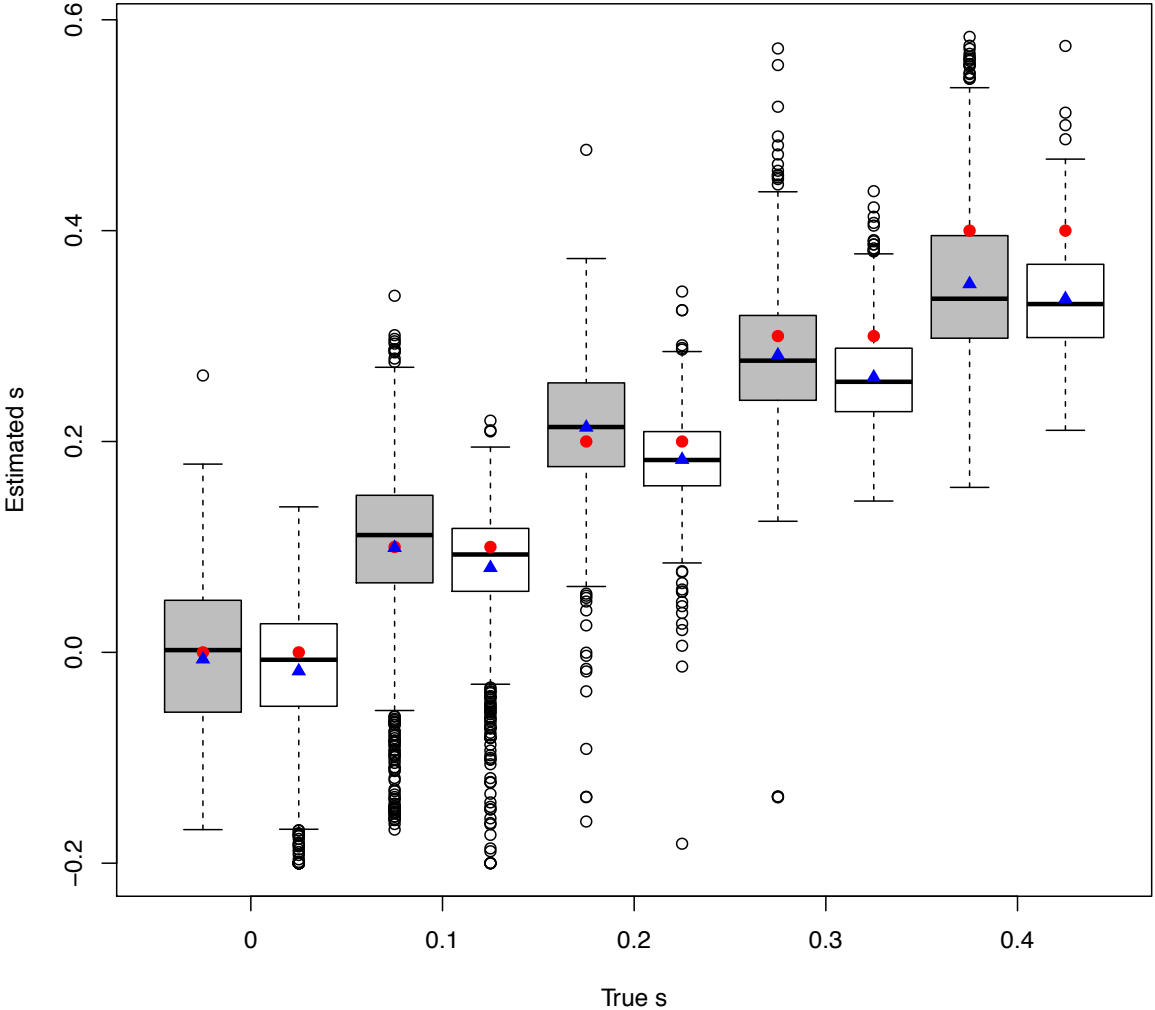

**Figure S9.** Boxplot for the estimated small selection coefficient from each simulation replicate of the Wright-Fisher diploid model with $N_e$=5000 simulated for 500 generations. The gray rectangles correspond to WFABC, and the white rectangles to the Malaspinas *et al.* (2012) method. The red circles are the true values for *s*, and the blue triangles are the mean of the estimated *s* values.

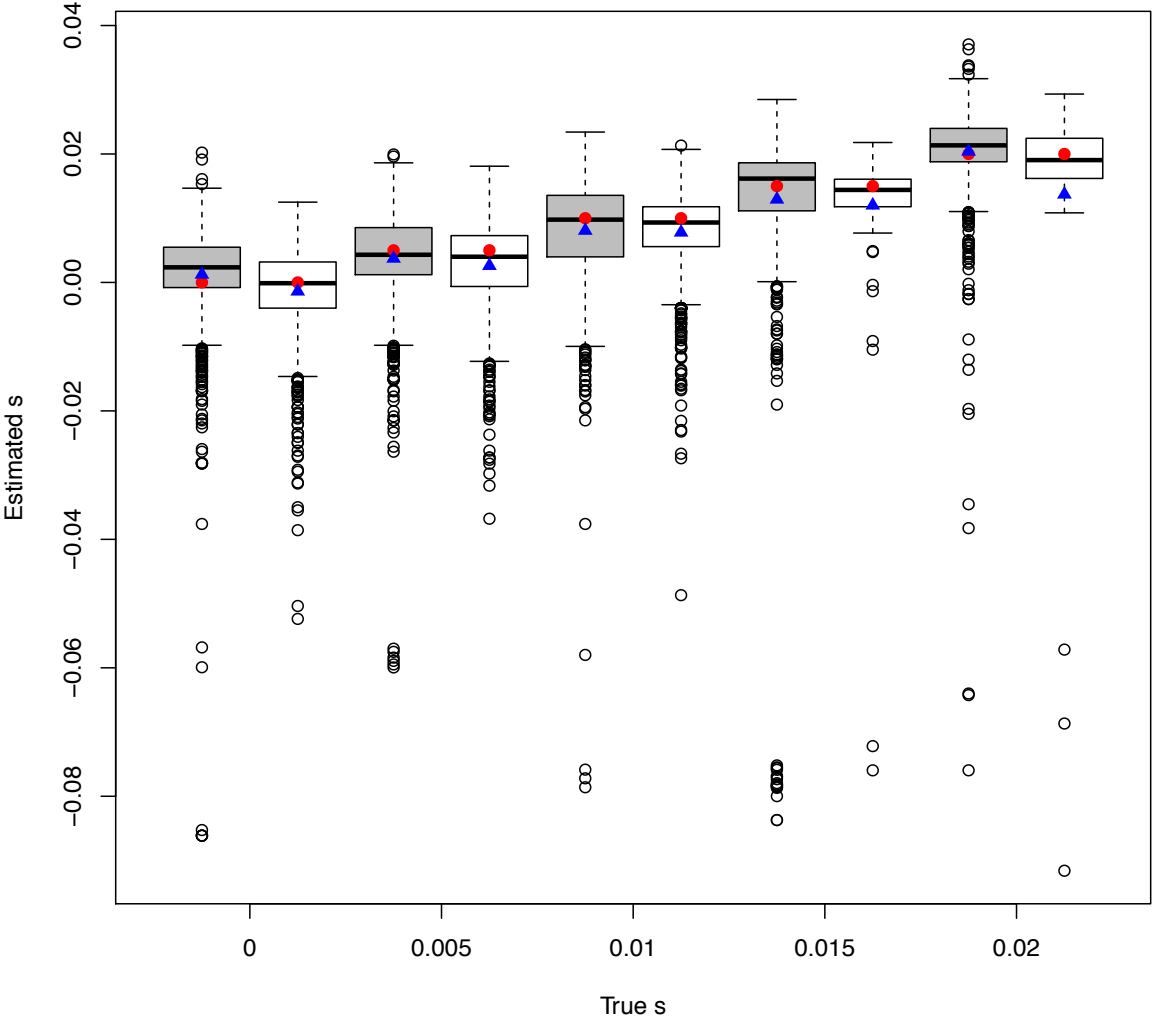

**Figure S10.** Boxplot for the estimated large selection coefficient from each simulation replicate of the Wright-Fisher diploid model with $N_e$=5000 simulated for 100 generations. The gray rectangles correspond to WFABC. The red circles are the true values for *s*, and the blue triangles are the mean of the estimated *s* values.

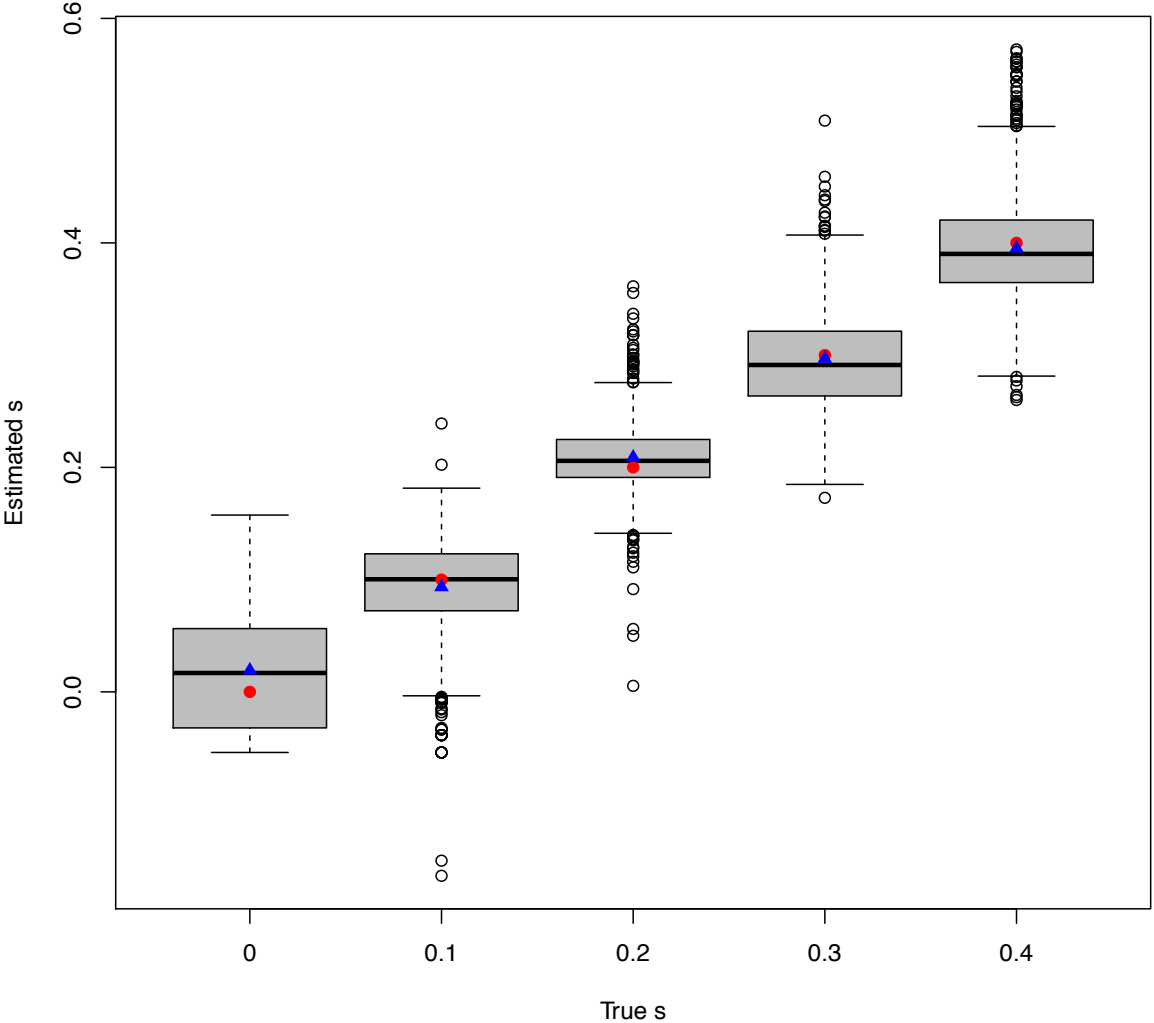

**Figure S11.** Boxplot for the estimated small selection coefficient from each simulation replicate of the Wright-Fisher haploid model with $N_e=1000$ simulated for 300 generations. The gray rectangles correspond to WFABC, and the white rectangles to the Malaspinas *et al.* (2012) method. The red circles are the true values for *s*, and the blue triangles are the mean of the estimated *s* values.

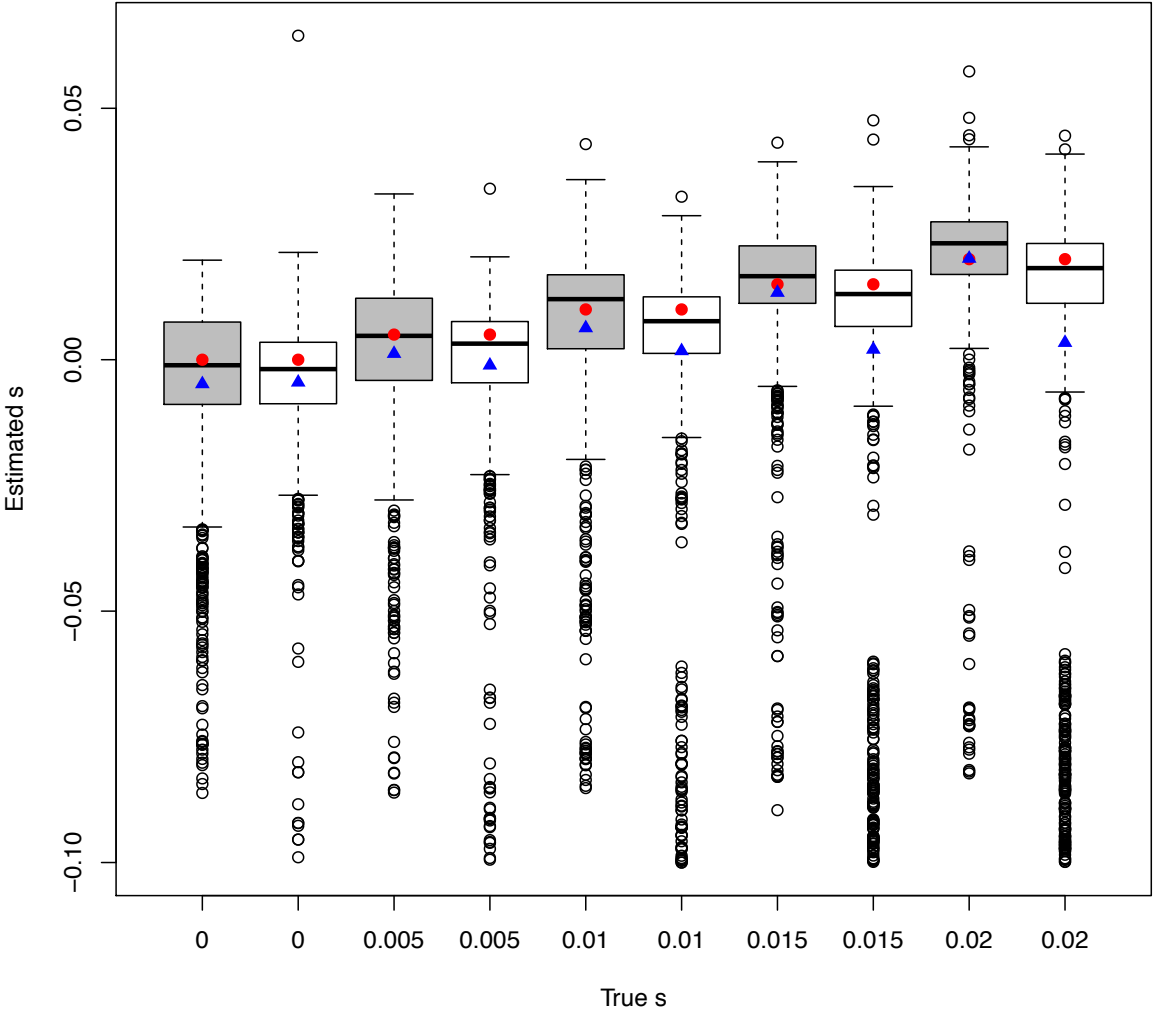

**Figure S12.** Boxplot for the estimated large selection coefficient from each simulation replicate of the Wright-Fisher haploid model with $N_e=1000$ simulated for 50 generations. The gray rectangles correspond to WFABC, and the white rectangles to the Malaspinas *et al.* (2012) method. The red circles are the true values for *s*, and the blue triangles are the mean of the estimated *s* values.

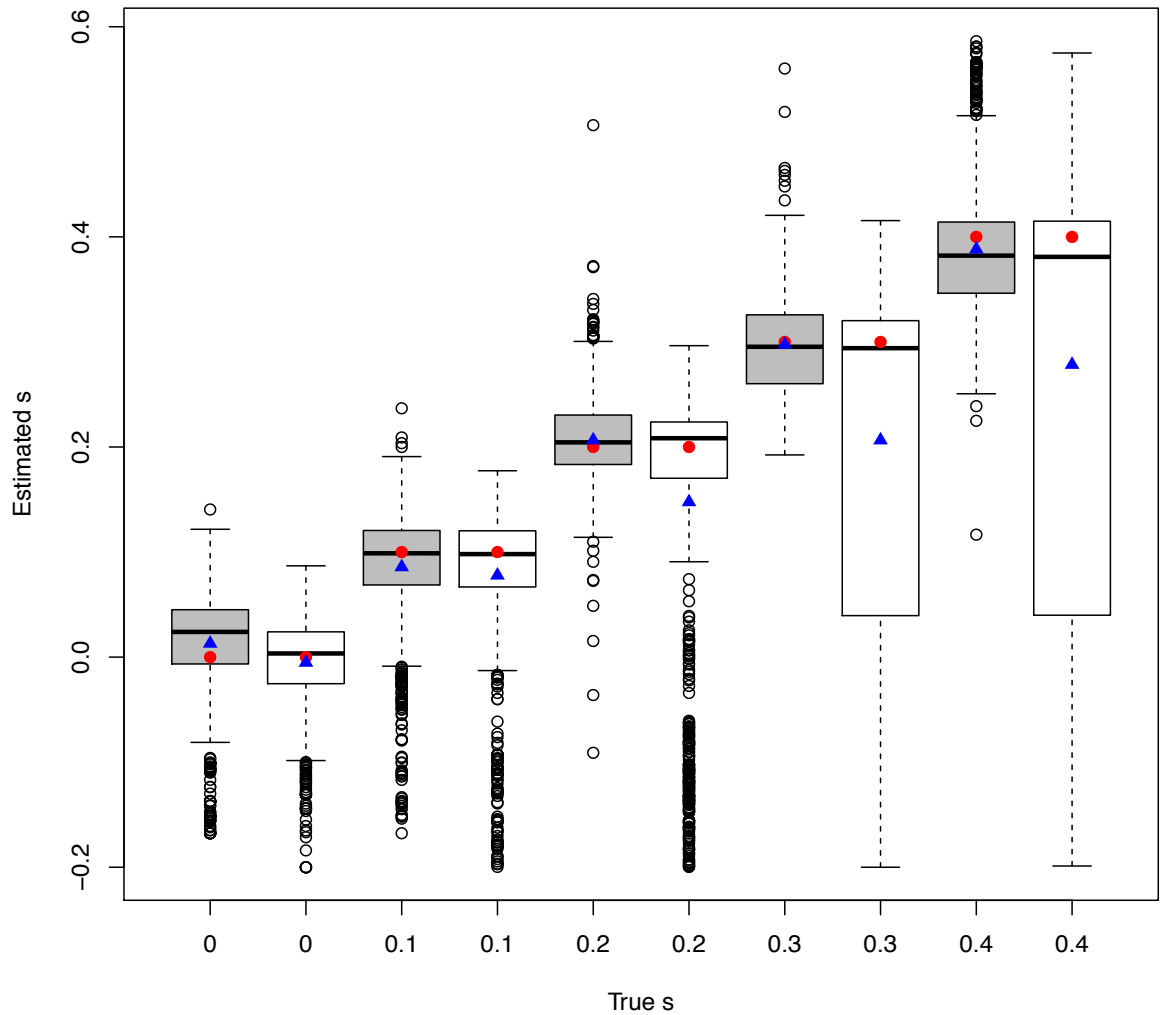

**Figure S13.** Two-dimensional joint posterior distribution for *s* and *h* for the moth *P. dominula* data using a uniform prior for *2N$_e$* between 100 and 10000. The grey lines delimit two-dimensional $\alpha$ highest posterior density regions for $\alpha$ =0.9 (largest region), 0.8, 0.7, 0.6, 0.5, 0.4, 0.3, 0.2 and 0.1 (smallest region).

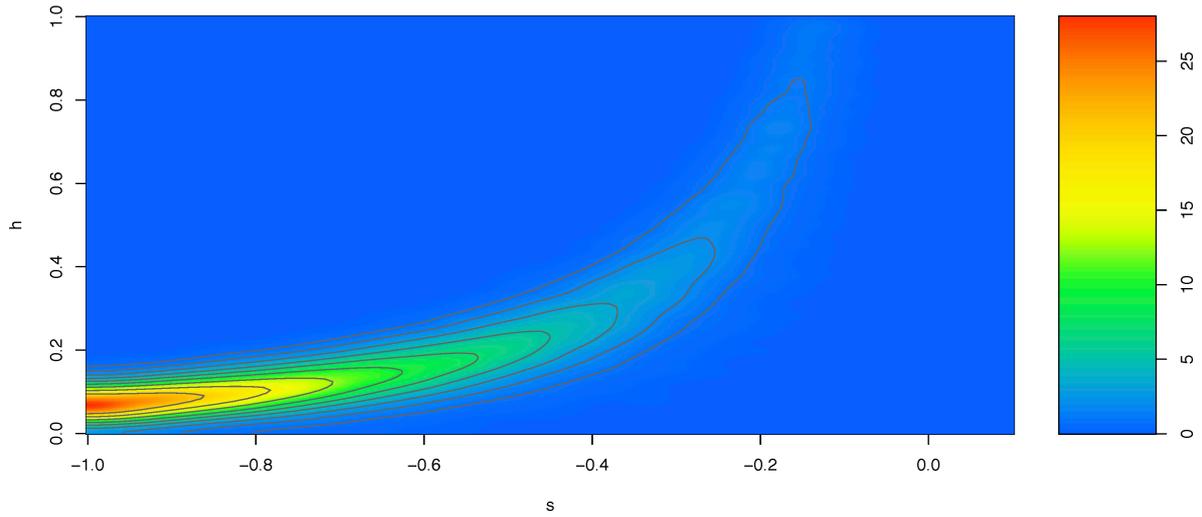

**Table S1.** RMSE and bias for the small *s* and $N_e$=200 scenario for the Wright-Fisher diploid model

|  |  | s=0 | s=0.005 | s=0.01 | s=0.015 | s=0.02 |
|---|---|---|---|---|---|---|
| Bias | WFABC | 0.0037 | 0.0039 | 0.0057 | 0.0060 | 0.0058 |
|  | Malaspinas *et al.* (2012) | -0.0064 | -0.0060 | -0.0026 | -0.0032 | -0.0018 |
| RMSE | WFABC | 0.031 | 0.032 | 0.032 | 0.030 | 0.031 |
|  | Malaspinas *et al.* (2012) | 0.026 | 0.027 | 0.028 | 0.027 | 0.028 |

**Table S2.** RMSE and bias for the big *s* and $N_e$=200 scenario for the Wright-Fisher diploid model

|  |  | s=0 | s=0.1 | s=0.2 | s=0.3 | s=0.4 |
|---|---|---|---|---|---|---|
| Bias | WFABC | -0.0065 | -0.00081 | 0.013 | -0.018 | -0.051 |
|  | Malaspinas *et al.* (2012) | -0.018 | -0.020 | -0.017 | -0.039 | -0.065 |
| RMSE | WFABC | 0.075 | 0.081 | 0.066 | 0.067 | 0.087 |
|  | Malaspinas *et al.* (2012) | 0.068 | 0.065 | 0.045 | 0.060 | 0.083 |

**Table S3.** RMSE and bias for the small *s* and $N_e$=5000 scenario for the Wright-Fisher diploid model

|  |  | s=0 | s=0.005 | s=0.01 | s=0.015 | s=0.02 |
|---|---|---|---|---|---|---|
| Bias | WFABC | 0.0012 | -0.0013 | -0.0019 | -0.0021 | 0.00040 |
|  | Malaspinas *et al.* (2012) | -0.0014 | -0.0024 | -0.0022 | - | - |
| RMSE | WFABC | 0.0094 | 0.0082 | 0.0091 | 0.014 | 0.0078 |
|  | Malaspinas *et al.* (2012) | 0.0073 | 0.0073 | 0.0069 | - | - |

**Table S4.** RMSE and bias for the big *s* and $N_e$=5000 scenario for the Wright-Fisher diploid model

|  |  | s=0 | s=0.1 | s=0.2 | s=0.3 | s=0.4 |
|---|---|---|---|---|---|---|
| Bias | WFABC | 0.019 | -0.0066 | 0.0090 | -0.0031 | -0.0049 |
|  | Malaspinas *et al.* (2012) | - | - | - | - | - |
| RMSE | WFABC | 0.052 | 0.044 | 0.034 | 0.045 | 0.053 |
|  | Malaspinas *et al.* (2012) | - | - | - | - | - |

**Table S5.** RMSE and bias for the small *s* and $N_e$=1000 scenario for the Wright-Fisher haploid model

|  |  | s=0 | s=0.005 | s=0.01 | s=0.015 | s=0.02 |
|---|---|---|---|---|---|---|
| Bias | WFABC | -0.0048 | -0.0038 | -0.0037 | -0.0017 | 0.00012 |
|  | Malaspinas *et al.* (2012) | -0.0045 | 0.0061 | -0.0083 | -0.013 | -0.017 |
| RMSE | WFABC | 0.019 | 0.017 | 0.020 | 0.018 | 0.016 |
|  | Malaspinas *et al.* (2012) | 0.015 | 0.018 | 0.025 | 0.035 | 0.041 |

**Table S6.** RMSE and bias for the big *s* and $N_e$=1000 scenario for the Wright-Fisher haploid model

|  |  | s=0 | s=0.1 | s=0.2 | s=0.3 | s=0.4 |
|---|---|---|---|---|---|---|
| Bias | WFABC | 0.013 | -0.014 | 0.0065 | -0.00247 | -0.012 |
|  | Malaspinas *et al.* (2012) | -0.0052 | -0.022 | -0.052 | -0.094 | -0.12 |
| RMSE | WFABC | 0.055 | 0.059 | 0.041 | 0.049 | 0.064 |
|  | Malaspinas *et al.* (2012) | 0.046 | 0.076 | 0.14 | 0.20 | 0.25 |